\RequirePackage{amsmath}
\documentclass{aa}

\pdfoutput=1
\usepackage{algorithm}
\usepackage{algpseudocode} 
\usepackage{graphicx}
\usepackage{txfonts}
\usepackage{mathrsfs}
\usepackage{natbib}
\usepackage{color}
\usepackage{soul}
\usepackage{footmisc}
\usepackage{multirow}
\usepackage{textcomp}

\def\Xv{\boldsymbol{X}}
\def\Zv{\boldsymbol{Z}}
\def\thetav{\boldsymbol{\theta}}
\def\epsilonv{\boldsymbol{\epsilon}}

\def\yv{\boldsymbol{y}}

\def\teff{T_{\text{eff}}}
\def\lzams{L_{\text{ZAMS}}}
\def\stage{t_{\star}}

\begin{document}

\title{Modelling the solar twin 18 Sco\thanks{Based on observations collected
  at the European Organisation for Astronomical Research in the
  Southern Hemisphere, Chile (run ID: 183.D-0729(A))}}
\author{M.~Bazot\inst{1,2}
\and O.~Creevey\inst{3}
\and J.~Christensen-Dalsgaard\inst{4}
\and J.~Mel\'endez\inst{5}
}

\institute{Center for Space Science, NYUAD Institute, New York University Abu Dhabi, PO Box 129188, Abu Dhabi, United Arab Emirates; mb6215@nyu.edu
\and
Division of Sciences, New York University Abu Dhabi, United Arab Emirates
\and
Universit\'e C\^ote d'Azur, Observatoire de la C\^ote d'Azur, CNRS, Laboratoire Lagrange, Bd de l'Observatoire, CS 34229, 06304 Nice cedex 4, France
\and
Stellar Astrophysics Centre, Department of Physics and Astronomy, Aarhus University, Ny Munkegade 120, DK-8000 Aarhus C, Denmark 
\and
Departamento de Astronomia do IAG/USP, Universidade de S\~ao Paulo, Rua do Mat\~ao 1226, S\~ao Paulo, 05508-900, SP, Brasil 
}
\abstract
    {Solar twins are objects of great interest in that they allow us to understand better how stellar evolution and structure are affected by variations of the stellar mass, age and chemical composition in the vicinity of the commonly accepted solar values.}
    {We aim to use the existing spectrophotometric, interferometric and asteroseismic data for the solar twin 18~Sco to constrain stellar evolution models. 18~Sco is the brightest solar twin and is a good benchmark for the study of solar twins. The goal is to obtain realistic estimates of its physical characteristics (mass, age, initial chemical composition, mixing-length parameter) and realistic associated uncertainties using stellar models.}
    {We set up a Bayesian model that relates the statistical properties of the data to the probability density of the stellar parameters. Special care is given to the modelling of the likelihood for the seismic data, using Gaussian mixture models. The probability densities of the stellar parameters are approximated numerically using an adaptive MCMC algorithm. From these approximate distributions we proceeded to a statistical analysis. We also performed the same exercise using local optimisation.
}
   {The precision on the mass is approximately 6\%. The precision reached on $X_0$ and $Z_0$ and the mixing-length parameter are respectively 6\%, 9\%, and 35\%. The posterior density for the age is bimodal, with modes at 4.67~Gyr and 6.95~Gyr, the first one being slightly more likely. We show that this bimodality is directly related to the structure of the seismic data. When asteroseismic data or interferometric data are excluded, we find significant losses of precision for the mass and the initial hydrogen-mass fraction. Our final estimates of the uncertainties from the Bayesian analysis are significantly larger than values inferred from local optimization. This also holds true for several estimates of the age encountered in the literature.}
   {}


\maketitle

\section{Introduction}

Amidst the labyrinthic zoology of stellar types and classes, one subset has gained considerable attention over the past few decades. These stars are called solar twins. Even though, with such a name, what they should be seems obvious, defining what they really are has so far been an ever-evolving process. Of course, one expects that a solar twin should have a physical state as close as possible to the Sun. This is actually the reason behind the relatively recent interest for these stars. They were first identified as a group by \citet{CdS81}, based on spectroscopic arguments. Therefore, one can immediately see that for a {\lq}good{\rq} solar twin to be classified as such will depend on the precision one can reach to estimate its physical properties. This also explains that, before this pioneering work, this class of stars has largely remained ignored. Being G2V objects, they are, on average, relatively faint, hence demanding large telescopes and high-resolution spectrographs in order to obtain a good precision on atmospheric parameters. Conversely, alongside the on-going improvement of spectroscopic instrumental methods, the threshold for classification as solar twin has evolved considerably and many observational campaigns have been carried over in order to detect these stars \citep{PdM97,King05,Melendez06,Melendez07,Takeda09,Melendez10,Datson12,Datson14,Melendez14b,PdM14,Ramirez14,Mahdi16}. As of today, roughly a hundred stars can be classified as solar twins, somewhat depending  on the exact criterion retained for classification.

Studying solar twins offers multiple perspectives. On a statistical level, they offer a good benchmark for solar-like populations. Some studies have focused on the properties of the Sun itself, trying to determine if it was an outlier with respect to some solar-twin samples \citep{Gustafsson98,Gustafsson08,Melendez14a,dosSantos16}. Exploiting further this idea, some other studies have explored potential planet-star connections using samples of solar twins. It was suggested for instance that planet-hosting stars were deficient in refractory elements\footnote{Elements with high condensation temperatures ($\gtrsim$ 900~K) and thus most likely to form rocky planets. Typical examples are Na, Mg, Al, Si, V, Cr,\dots}. This is also true for the Sun itself, which is refractory-deficient with respect to most of the known solar twins \citep{Ramirez10}. They were also used to study the problem of Li depletion in planet-hosting stars \citep{Israelian09}. Studying a sample of \emph{approximate twins} (in the sense that their ages span a range that encompasses largely the solar age), \citet{Baumann10} claimed that this trend is purely evolutionary and is not correlated to the presence of planets \citep[see also][]{Monroe13,Carlos16}. 

From a stellar-modelling point of view, solar twins are extremely interesting targets. We first recall that stellar evolution codes were largely developed using the Sun as a reference observational benchmark. Indeed it is the only star for which we have extremely precise measurements, independent of modelling, for age \citep{Bouvier10} and mass \citep{Olive14}. Furthermore, the observational precision on its radius \citep{Emilio12} and luminosity \citep{Frohlich04} allows to calibrate solar models and assess their accuracy through helioseismic inversion \citep[see e.g.,][]{Thompson91,Basu97,JCD02,Basu16}. 

As a result, we know with excellent confidence the main physical processes at play in  the solar interior. However, these solar models are usually built on assumptions such as spherical symmetry, neglecting rotation or magnetic fields, that may quickly break down when the mass of the star varies. Likewise, the sizes of convective envelopes or cores depend strongly on the stellar mass and evolutionary phase. For low-mass stars, the outer convective envelope becomes deep enough that the radiative core disappears. This may change the stability properties of the star against small perturbations \citep{Gabriel64,Gabriel67,RL12}. For masses larger than the Sun, the convective envelope rapidly becomes very thin while convective cores start to appear. Their modelling is somewhat uncertain. This can be due to numerical issues in treating simultaneously microscopic diffusion and nuclear reactions in a convective region \citep[see for instance][]{JCD08a}. It might also be the consequence of other phenomena such as double-diffusive convection, in that case taking the form of semiconvection \citep{Moore16}.

A good property of solar twins is that they shall not enter such regimes. At the same time, they might differ slightly from the Sun. Hence, by studying them, we can be confident that the general assumptions made for our stellar models hold. But at the same time, we can test this model by letting the stellar structure vary.

In this paper, we focus on 18 Sco, the brightest solar twin. It was the first observed by asteroseismology \citep{Bazot11,Bazot12}. This allowed to estimate its global seismic properties such as the large separation, but also individual frequencies. In this paper we want to assess the impact of such measurements on our knowledge of the physical state of 18 Sco. In particular, we want to obtain statistically robust estimates of its global physical parameters but also of the related uncertainties. 

In Sect.~\ref{sect:statmod} we present the statistical model we use to estimate these parameters. We also present the two estimation methods, namely Bayesian probability density estimation and frequentist optimisation, that we selected to carry out the estimation. Importantly, we reassess the seismic data given in \citet{Bazot12}, which needs to be treated carefully before being incorporated in a statistical model. In Sect.~\ref{sect:blabla} we present our results and discuss them.

\section{Statistical model}\label{sect:statmod}

\begin{figure*}[t!]
\center
\includegraphics[width=0.475\textwidth]{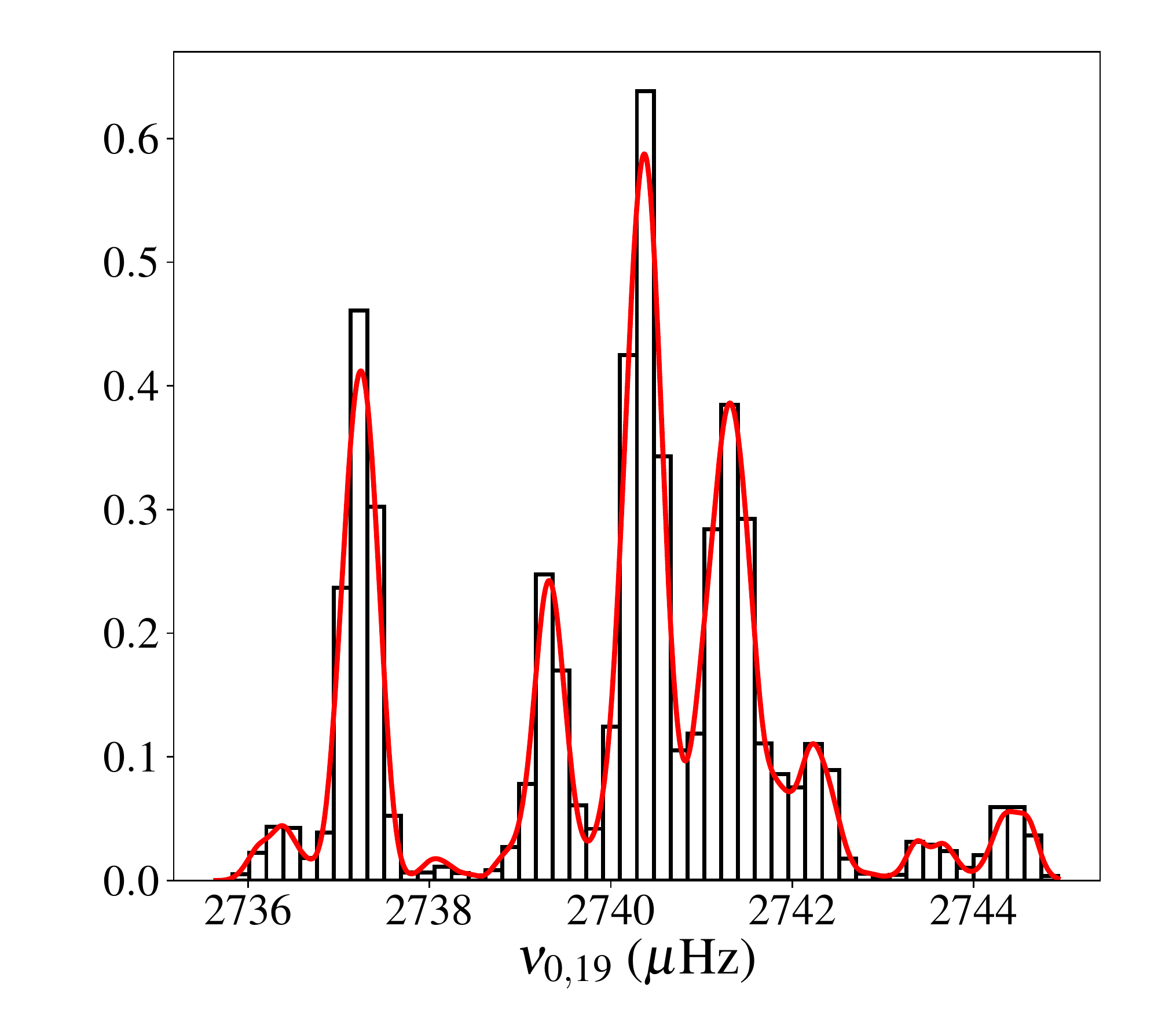}
\includegraphics[width=0.475\textwidth]{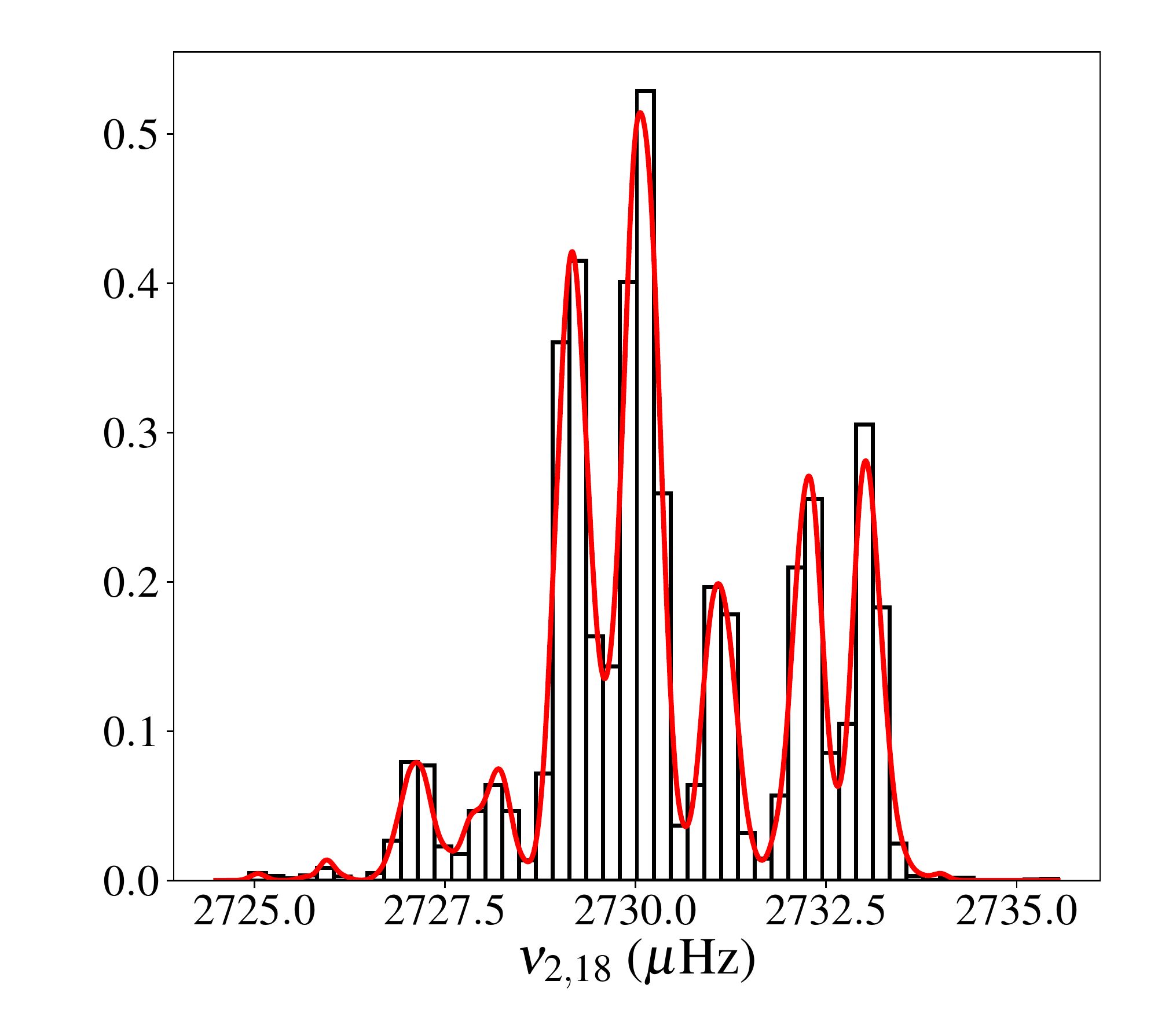}
\caption{Examples of problematic PDFs of individual frequencies for $n = 19$, $l = 0$ (left panel) and $n = 18$, $l = 2$ (right panel). The black lines represent the histogram estimate and the red line a kernel-density estimation. }
\label{fig:nu_dist}
\end{figure*}

\subsection{Generic formulation}

By statistical model, we mean a mathematical formulation of the behaviour of our observations. A simple approach is to consider an additive model made of a deterministic and a random part, and hence having the form
\begin{equation} 
\Xv = \mathcal{S}(\thetav) + \epsilonv,
\end{equation} 
where $\Xv$ are the observations, $\mathcal{S}(\thetav)$ is the deterministic part (being a function of $\thetav$, the stellar parameters we want to estimate) and $\epsilonv$ the realisation of a random variable that represents the observational noise.

In the following, we will make the (somewhat optimistic) assumption that our model is unbiased, that is that the noise component has zero mean and hence that the stellar model $\mathcal{S}$ is the expectation value of our observations. We assume that the observations are independent. The likelihood has thus the form
\begin{equation}
\pi(\Xv|\thetav) = \prod_i \pi(X_i|\thetav),
\end{equation}
where the $X_i$ are the components of the vector $\Xv$.
\subsection{Data and the physical model}
\label{sect:dataandphysicalmodel}
\subsubsection{Non-seismic data}

\begin{table}
\caption{Recent atmospheric parameters found in the literature and weighted average used in this work.}
\label{tab:atmos}      
\centering                                      
\begin{tabular}{l c c c}          
\hline\hline
\\[-3.mm]                                   
$\teff$ (K)& $\log g$& [Fe/H] & Reference\\
\hline
\\[-3.mm]
 $5823\pm6$  & $4.45\pm0.02  $  & $0.054\pm0.005$& \citet{Melendez14b}\\
 $5818\pm3$  & $4.457\pm0.010$  & $0.054\pm0.004$& \citet{Nissen15}\\
 $5809\pm6$  & $4.434\pm0.012$  & $0.046\pm0.006$& \citet{Spina16}\\
 $5817 \pm 4$& $4.448 \pm 0.012$  & $0.052 \pm 0.005$& This work\\
\hline
\end{tabular}
\end{table}

One of the main advantages of studying solar twins is that stellar spectra can be analysed differentially relative to the solar one. The departures from the latter are small enough that they can be treated as a first-order perturbations. This assumes that a stellar spectrum and a solar one have been obtain from the same instrument \citep{Melendez14b}. This translates in turn into smaller uncertainties on the atmospheric parameters  $\teff$, $\log g$ and [Fe/H] than those usually quoted for other stars \citep[see e.g.,][]{Melendez14b,Nissen15,Spina16}.

In this work we considered three recent estimates of the atmospheric parameters of 18 Sco from \citet{Melendez14b}, \citet{Nissen15} and \citet{Spina16}. They are given in Table~\ref{tab:atmos}. Assuming that data points have Gaussian parent distributions with the same mean but different standard deviations, a reasonable new estimate of these atmospheric parameters is the weighted average of the sample. The variance of the weighted mean was then used to compute the associated uncertainties. The resulting parameters are $\teff = 5817 \pm 4$~K, $\log g = 4.448 \pm 0.012$, [Fe/H] = 0.052 $\pm$ 0.005. For modelling purposes, we used only the effective temperature and surface metallicity, choosing to constrain our model with the luminosity rather than $\log g$. The conversion between [Fe/H] and $Z/X$, which we effectively use as the output of the stellar code, was performed using the solar ratio $(Z/X)_{\odot}$ from \citet{Grevesse98}. We note that different values for this ratio have been derived since then \citep{Asplund05}. However, these lower estimates raised the, still unsolved, solar-abundance problem, which put solar models and solar observations at odds \citep[see e.g.,][]{Guzik06,Castro07,Basu08,Antia11,Gough12,Basu16}. For the sake of conciseness and simplicity we chose to overlook this issue in the present study.
 
For the luminosity, we selected the value given by \citet{Boyajian13}. It is based on an estimate of the bolometric flux $F_b = (17.34 \pm 0.09)\times 10^{-8}$ erg s$^{-1}$cm$^{-2}$. This was obtained using an aggregate of various photometric fluxes and compared to a library of stellar spectra \citep{Pickles98}.  The corresponding luminosity is $ L = 1.0438 \pm 0.0120$~L$_{\odot}$.

A natural choice for the radius is the one derived by \citet{Bazot12}. It was obtained using the CHARA interferometric array and the PAVO interferometer. Its value is $1.010 \pm 0.009$~R$_{\odot}$. 

The luminosity and radius were obtained using the \emph{Hipparcos} parallax \citep{vanLeeuwen07}. We note that a much higher value of $1.166 \pm 0.026$~R$_{\odot}$ has been derived by \citet{Boyajian12}. Their subsequent modelling leads however to far too large ages for 18 Sco. Consequently we decided not to take it into account through averaging (as we did for the effective temperature). Some recent analyses of angular diameters derived  by \citet{Casagrande14} and \citet{White18} claim some possible systematic errors for stars observed by \citet{Boyajian12} that are not very well resolved, justifying our decision not to use the radius measurement.
Finally, note that using the independently-determined, but strongly-correlated observations effective temperature, radius and luminosity provides consistency check between the existing constraints. It is indeed not always easy to find a model that reproduce them all, as shown by the case of $\alpha$~Cen~A \citep{Miglio05}. \\

For the non-seismic data, we always choose Gaussian densities. Therefore, the likelihood for these observations is proportional to
\begin{equation}\label{eq:ns_likelihood}
\exp \left[ -\frac{1}{2}\sum_{i=1}^{n_{\mathrm{ns}}} \frac{(X_i - \mathcal{S}_i(\thetav))^2)}{\sigma_i^2} \right],
\end{equation}
where $i$ labels the non-seismic observations and $n_{\mathrm{ns}}$ is their number.

\begin{figure}
\center
\includegraphics[width=0.475\textwidth]{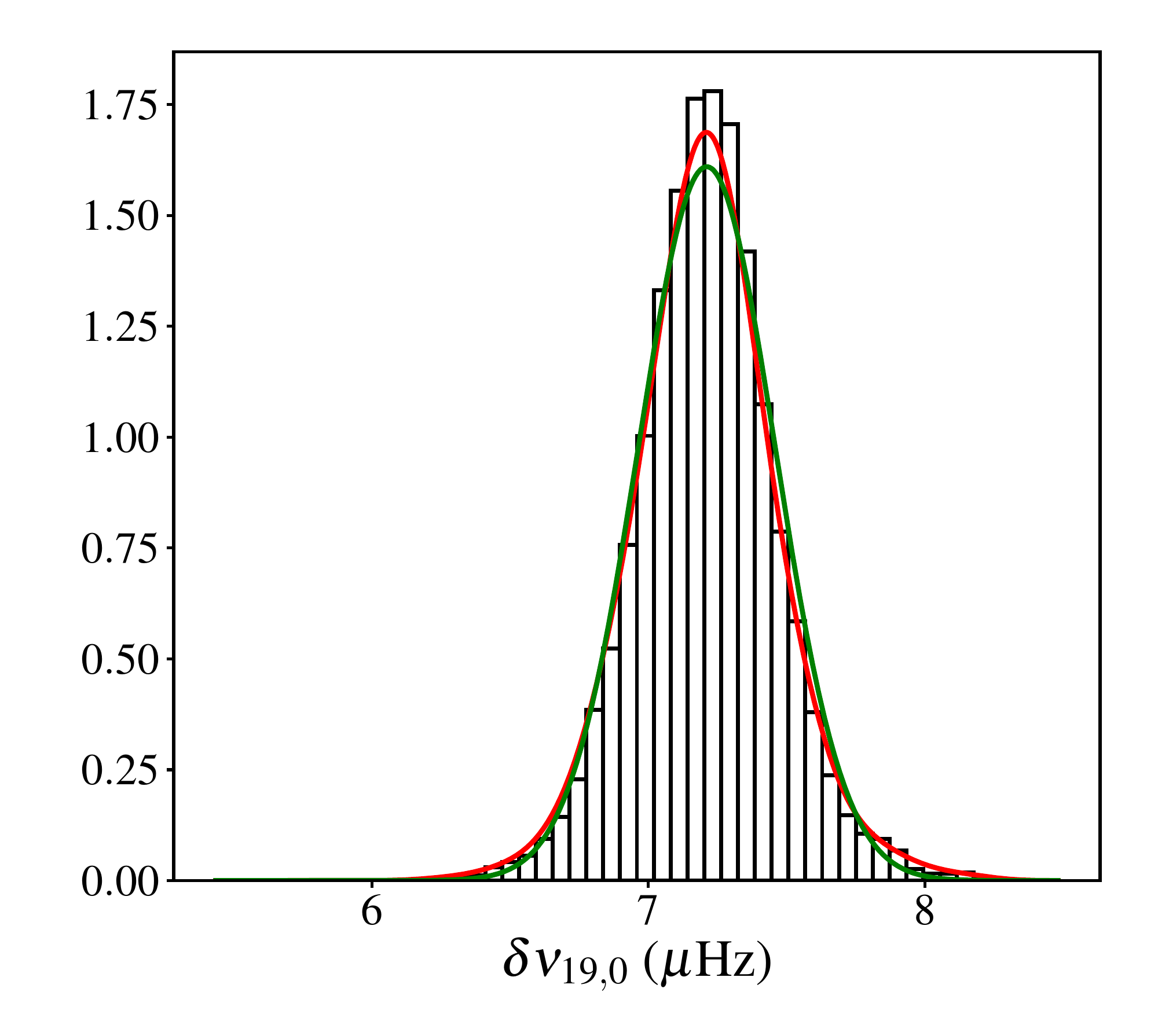}
\caption{PDF for the small separation $\delta\nu_{0,19}$. The black lines represent the histogram estimation, the red line the kernel-density estimation and the green one the result of mixture model fitting.}
\label{fig:ss_dist}
\end{figure}

\subsubsection{A reassessment of the seismic data}\label{sect:sismo}

The frequencies provided in Table~3 of \citet{Bazot12} could potentially be used to derive seismic indicators such as the small separations (see below). However, one needs to clearly understand what these estimates mean. In order to do so, one must go back to the output of the Markov chain Monte Carlo (MCMC) simulation that was used to estimate the oscillation frequencies of 18~Sco. A thorough examination indicates that the marginal posterior density functions (PDFs) for the $\nu_{n,l}$ are extremely complex, mostly reflecting the noisy nature of the data and the difficult spectral window induced by the ground-based observations. Typical examples of such distributions are shown in Fig.~\ref{fig:nu_dist}. Therefore, even though the estimates given in \citet{Bazot12} provide quantitative assessments on the oscillation frequencies, they are difficult to use in statistical models as such, that is to provide a likelihood for asteroseismic diagnostic. They are statistical summaries that only very partially capture the shape of their underlying parent distribution.

 We chose not to derive the seismic indicators based on the estimates of the individual frequencies given in \citet{Bazot12}. This is motivated on one hand by the potential difficulty there is when facing likelihoods with multiple modes\footnote{Not to be mistaken with an oscillation eigenmode, obtained from the pulsation equation. Here a mode is used in its statistical sense, i.e. the local maximum of a probability density.}, such as shown in Fig~\ref{fig:nu_dist}. These could lead to many degeneracies in the solution to the estimation problem. On the other hand, dealing with individual frequencies demands to take into account surface effects \citep[see e.g.][]{Kjeldsen08}. As a general rule, this could be problematic for any given star \citep{Bazot13}.


A much more robust approach consists in analysing directly the posterior probability density of these seismic indicators themselves. This is extremely straightforward since one simply has to combine the MCMC samples for the individual frequencies to obtain the sample of a frequency separation or ratio. For instance, the two samples $\{ \nu_{n,l}^{(1)},\dots,\nu_{n,l}^{(N)}\} \sim \pi(\nu_{n,l}|\yv)$ (where $\yv$ stands for the observed time series) and $\{ \nu_{n-1,l+2}^{(1)},\dots,\nu_{n-1,l+2}^{(N)}\} \sim \pi(\nu_{n-1,l+2}|\yv)$ allow us to obtain the sample $\{ \nu_{n,l}^{(1)} - \nu_{n-1,l+2}^{(1)},\dots,\nu_{n,l}^{(N)} - \nu_{n-1,l+2}^{(1)}\} \sim \pi(\delta\nu_{n,l}|\yv)$. Here we used the definition $\delta\nu_{n,l} = \nu_{n,l} - \delta\nu_{n-1,l+2}$ for the small separation. These were chosen because they are mostly sensitive to the innermost regions of the star \citep{Tassoul80,Roxburgh94}.

It turns out that these samples are much easier to study than those of individual frequencies. An example of such a situation is shown in Fig.~\ref{fig:ss_dist}. A comparison with Fig.~\ref{fig:nu_dist} shows that most of the multiple maxima found in the marginal PDFs for $\nu_{19,0}$ and $\nu_{18,2}$ are not found in the corresponding PDF for $\delta\nu_{19,0} = \nu_{19,0} - \nu_{18,2}$. This is easy to understand if one considers that from one iteration of the MCMC algorithm to the other the value of these frequencies might vary greatly. However, their average small separation will have to remain centred around the same value in order to reproduce the seismic data.

This first step considerably simplifies our analysis. We also stress that any subsequent study of 18~Sco using the HARPS data should adopt this approach. However, our marginal PDFs for the small separations remain complicated. We noticed that some of them depart from normal densities. In most cases, they display important asymmetries and, sometimes, multiple modes. A very interesting approach in order to manipulate these distributions is to model them as mixtures of normal distributions \citep[see e.g.][]{FS06}. Formally if we consider a random vector $\Zv$ with density $p_{\Zv}$, we can model the latter as follows
\begin{equation}\label{eq:mixtnorm}
\displaystyle
p_{\Zv}(\Zv) = \sum_{j=1}^M P_j\mathcal{N}(\boldsymbol{\mu_j}, \boldsymbol{\Sigma_j}), 
\end{equation}
with $\mathcal{N}(\boldsymbol{\mu}_j, \boldsymbol{\Sigma}_j)$ a multivariate normal distribution with mean $\boldsymbol{\mu_j}$, a vector of size $N = \dim(\Zv)$, and covariance matrix $\boldsymbol{\Sigma_j}$, of size $N\times N$, and subject to the constraints
\begin{equation}\label{eq:mixtfact}
\displaystyle
\sum_{j=1}^M P_j = 1 \text{ and } P_j \geq 0.
\end{equation}

Equation (\ref{eq:mixtnorm}) gives a Gaussian mixture model for a vector. Ideally, this is how one would treat any seismic indicator used to constrain a stellar model. For instance, when dealing with the so-called frequency ratios $r_{01}$ or $r_{10}$ \citep{Roxburgh03} one has to take into account that any values evaluated for order $n_1$ and $n_2$, such that $n_1 \neq n_2$, are correlated through the frequencies that enter their computations \citep[see e.g.][]{SA13}. This is not true, however, of the individual small separations which are uncorrelated for different values of the couple $(n,l)$. Lets call $\Xv^{\mathrm{sis}}$ that vector that regroups all the seismic indicators we wish to reproduce. In the case its components are only small separations, we can model each one with a separate Gaussian mixture model. This is a very convenient simplification. Indeed, a popular approach to estimate the parameters of a mixture model is the expectation-minimisation (EM) algorithm. As it turns out estimating the parameters of a Gaussian mixture model of the form (\ref{eq:mixtnorm}) using an EM algorithm becomes increasingly difficult when the dimension of the problem increases. Using small separation allows us to bypass this technical difficulty. We could then use a separate Gaussian mixture model to model each individual small-separation likelihood

\begin{equation}\label{eq:comp_likelihood}
\displaystyle
\pi(\delta\nu_{n,l}|\theta) = \sum_{j=1}^{M}P_{j}\mathcal{N}(\mu_{j},\sigma_{j}^2).
\end{equation}

If we map (bijectively) the $(n,l)$ couples for which we have measured a small separation onto a single index $i$ then the seismic data vector can be written $\Xv^{\mathrm{sis}} = (\delta\nu_1,\dots,\delta\nu_i,\dots,\delta\nu_N)$. Here, $N$ is the number of observed small separations. The corresponding likelihood is then
\begin{equation}\label{eq:likelihood}
\displaystyle
\pi(\Xv^{\mathrm{sis}}|\theta) = \prod_{i=1}^N\left[\sum_{j=1}^{M_i}P_{j,i}\mathcal{N}(\mu_{j,i},\sigma_{j,i}^2)\right].
\end{equation}
The parameters of the Gaussian mixtures were estimated separately, for a given $i$, following the simple version of the EM algorithm given in \citet[]{Bishop95}. In Tables~\ref{tab:mixt_ss} we give the main characteristics of the mixture model we used to describe the small separations. 

\onltab{2}{
\begin{table}
\caption{Mixture models parameters for the small separations.}              
\label{tab:mixt_ss}      
\centering                                      
\begin{tabular}{ccccc}          
\hline\hline                        
$l$ & $n$&  Mean & Standard deviation & Weight\\    
\hline
\\[-3.mm]                                   
0 &	14 &  12.23	& 2.86	& 1.	\\[0.8mm]
0 &	15 &  10.32	& 2.49	& 1.	\\[0.8mm]
0 &	16 &  10.29	& 3.09	& 1.	\\[0.8mm]
0 &	18 &  9.58	& 2.90	& 1.	\\[0.8mm]
0 &	19 &  7.21	& 0.25	& 1.	\\[0.8mm]
0 &	22 &  10.66	& 1.01	& 0.63	\\[0.8mm]
  &	   &  13.96	& 0.89	& 0.14	\\[0.8mm]
  &	   &  9.15	& 0.64	& 0.23	\\[0.8mm]
0 &	24 &  16.11	& 1.12	& 0.66	\\[0.8mm]
  &	   &  7.76	& 2.87	& 0.34	\\[0.8mm]
0 &	25 &  8.47	& 3.69	& 0.30	\\[0.8mm]
  &	   &  6.64	& 1.14	& 0.70	\\[0.8mm]
0 &	26 &  20.88	& 1.98	& 0.43	\\[0.8mm]
  &	   &  17.11	& 3.08	& 0.57	\\[0.8mm]
1 &	15 &  22.64	& 1.00	& 0.15	\\[0.8mm]
  &	   &  20.87	& 1.44	& 0.65	\\[0.8mm]
  &	   &  16.46	& 1.91	& 0.20	\\[0.8mm]
1 &	16 &  17.88	& 0.73	& 0.72	\\[0.8mm]
  &	   &  14.75	& 2.04	& 0.05	\\[0.8mm]
  &	   &  20.78	& 0.81	& 0.23	\\[0.8mm]
1 &	17 &  17.45	& 1.52	& 0.35	\\[0.8mm]
  &	   &  17.94	& 0.57	& 0.34	\\[0.8mm]
  &	   &  17.71	& 3.31	& 0.31	\\[0.8mm]
1 &	18 &  16.13	& 0.58	& 1.	\\[0.8mm]
1 &	19 &  19.11	& 1.09	& 0.06	\\[0.8mm]
  &	   &  12.19	& 1.27	& 0.94	\\[0.8mm]
1 &	20 &  18.46	& 2.22	& 0.24	\\[0.8mm]
  &	   &  16.07	& 0.43	& 0.76	\\[0.8mm]
1 &	21 &  16.32	& 0.76	& 1.	\\[0.8mm]
1 &	23 &  16.71	& 2.72	& 1.	\\[0.8mm]
1 &	24 &  15.68	& 1.55	& 0.13	\\[0.8mm]
  &	   &  9.25	& 2.02	& 0.87	\\[0.8mm] 
\hline                                             
\end{tabular}
\end{table}

}

A mixture model has the advantage, in the framework of MCMC sampling, of being easy to compute, since it has a simple analytic form. However, a word of caution is in order concerning the current implementation of this methodology in our analysis. One should note that we are using a rudimentary approach to mixture model fitting. In particular, we evaluate, by visual inspection, the number $M$ of normal distributions to be included in the sum on the right-hand side of Eq.~(\ref{eq:mixtnorm}).


\subsubsection{ASTEC}\label{sect:ASTEC}

Our model for the evolution and oscillations of the stellar structure, $\mathcal{S}$, is composed of the Aarhus STellar Evolution Code (ASTEC) and {\tt adipls}. Both have been extensively described in the literature \citep{JCD82b,JCD08a,JCD08b}. Here we simply state the main settings we adopted.

We assumed a non-rotating, non-magnetic star. The opacities and equation-of-state tables in which we interpolate are taken from the OPAL collaboration, respectively from \citet{Iglesias96} and \citet{OPAL02}. Nuclear reaction rates were taken from \citet{Angulo99} with the additional inclusion of the values obtained by the LUNA collaboration for the $^{14}$N(p,$\gamma$)$^{15}$O reaction \citep{Formicola04}. Diffusion was included for He and heavy elements. These latter are treated as a block. It is fine to do so with ASTEC as long as we do not try to model stars with convective cores \citep{JCD08a}.

Many parameters can be tuned in ASTEC. We only let a small subset vary, namely the mass, $M$, the age, $\stage$, the initial metallicity and hydrogen mass fraction, $Z_0$ and $X_0$. We also have the mixing-length parameter $\alpha$ as a free parameter. The latter is the proportionality coefficient between the mean-free path of a fluid element, in the mixing-length picture as described by \citet{BV58}, and the pressure scale height. Therefore, $\mathcal{S}$ is a mapping from a subspace of the parameter space to the observation space in which the vectors are respectively $\thetav = (M,\stage,X_0,Z_0,\alpha)$ and $\Xv = (\teff, L, Z/X, R, \{\nu_{n,l}\})$.

\subsection{Bayesian estimation}

\subsubsection{Bayesian statistical model}\label{sect:bsm}

Bayesian density estimation is one of the two strategies we adopt to obtain values for the parameters of 18 Sco. To that effect, we shall supplement our parametric statistical model with a prior density for the parameters as per Bayes' formula
\begin{equation}\label{eq:bayes}
\pi(\thetav |\Xv) \propto \pi(\thetav)\pi(\Xv|\thetav).
\end{equation}
Here $\pi(\Xv|\thetav)$ is the likelihood, which is in our case computed as the product of Eqs.~(\ref{eq:ns_likelihood}) and (\ref{eq:likelihood}). $\pi(\thetav|\Xv)$ is the Posterior Density Function of the parameters conditional on the data, $\Xv$, and $\pi(\thetav)$ is the aforementioned prior density. The former is the object we are ultimately interested in, since its knowledge allows us to use the tools of statistics to provide estimates of the parameters. The latter is the fundamental feature on which Bayesian statistics rely and thanks to which one can, in practice, switch from the observations to the parameters being the random quantities in the problem.

The prior density encodes the information we possess on the parameter before carrying over the estimation. Its use has been the subject of many discussions and debates for many decades. These are far outside the scope of this paper. For the present study, suffice to say that one always needs to specify carefully the prior density considered. Indeed, its formulation will condition the final outcome of the estimation process. As general rule, using two different priors $\pi_1(\thetav)$ and $\pi_2(\thetav)$ in (\ref{eq:bayes}) shall ultimately result in different a posteriori estimates $\widehat{\thetav}_1$ and $\widehat{\thetav}_2$. That being said this does not mean that Bayesian Statistics are more {\lq}subjective{\rq} than a frequentist method. They simply provide a way to formalise assumptions one may have on the outcome of the estimation process, for instance forbidden regions in the parameter space or previous independent measurement on some of the parameters. In stellar physics a typical example are stellar mass estimates for members of close binaries \citep[see for instance][]{Bazot16}.

The prior information on the stellar parameters is sparse. In this study we only used uniform densities, their properties are given in Table~\ref{tab:priors}. The only parameter for which reliable prior measurements exist is the mass. A previous study by \citet{Bazot12} gives an estimate of the mass, namely $1.02 \pm 0.03$~M$_{\odot}$. This was obtained by combining density and radius estimates through homology relations. Therefore, this estimate is based on the same data we are using here and a much cruder physical model than ASTEC. For these reasons we decided not to use it as a prior, but simply compare it to the results inferred from the PDF. 


  Regarding the other stellar parameters, only two of them have clear cut upper limits. The age of 18 Sco ought to be smaller than $\sim$13~Gyr, the age of the Universe\footnote{A more precise value is provided by WMAP: $13.772\pm0.059$~Gyr \citep{Planck16}. Such a level of precision is not required here. In all the samples from our MCMC simulations, the model with the largest age is about 12.6~Gyr.}. For $X_0$ helium measurements have shown that the earliest galaxies have an helium mass fraction $\leq 0.25$ 
\citep{Olive04,Aver13}.
If we neglect the metal abundance in these very old galaxies at the epoch of their formation, we can set an upper bound on the initial hydrogen mass fraction, $X_0 = 0.75$.
  
The other parameters are less-well constrained. In practice, upper and lower bounds can be obtained using test runs of our MCMC algorithm (see Sect.~\ref{sect:MCMC} below). After analysing their outcome, we can redefine the domain of definition of our prior by excluding regions of the space of parameters in which we are confident that models will not be accepted by the algorithm. This empirical procedure using MCMC test runs was used to set the priors for $Z_0$ and $\alpha$. The resulting domains are sometimes very large with respect to the region in which the marginal PDF significantly differs from zero. This is largely due to the fact that if a parameter is only allowed to vary over a narrow region, then the approximate PDF could be artificially increased close to the boundaries. This is due to the MCMC algorithm trying to go past the upper or lower limits and thus getting {\lq}swamped{\rq} near the boundaries. Therefore, only sound physical arguments, such as the ones given for the upper bounds for $\stage$ and $X_0$, shall motivate strong restrictions in the space of parameters, and we chose to err on the safe side for the numerical setup. 

  One may wish to include an $Y_0 - Z_0$ relation in the prior. These have been observed previously. Some studies have shown, for instance, that there exists a linear relation between galactic abundances of helium and metals \citep{Izotov04,Fukugita06,Balser06,Casagrande07}, and hence between $X_0$ and $Z_0$. Such relations have been used previously for stellar modelling \cite[see e.g.][]{Deal17}. However, we do not wish to incorporate a priori correlations between the parameters but would rather study them a posteriori. Thus we retain independent uniform priors for both $Z_0$ and $X_0$.

  Finally, the case of $\alpha$ is a difficult one. The current formulation of the mixing length is somewhat heuristic, adopted in order to provide a convective flux in one-dimensional models. Numerical simulations far more precise than the one used here exist and have shown the mixing-length parameter to remain fairly constant across the HR diagram, at least for solar analogues \citep{Trampedach14}. So far, these simulations have not been used directly to fit stellar observations. Some hybrid one-dimensional stellar codes that interpolate in the tables obtained from three-dimensional simulations have developed in order to obtain solutions in the upper stellar layers, and in particular the superadiabatic layers \citep{Sonoi15,Ball16,Houdek17,Joergensen17}. These were mostly developed to account for surface effects on oscillation frequencies but could be of great interest to simply provide a more robust formulation of surface convection. Nevertheless, we do not have such a code implemented together with an MCMC interface. Moreover, the testing required to ensure good performance of a stellar code interpolating in a grid of three-dimensional atmosphere in the context of Bayesian estimation is outside the scope of this study. This implies that we have to let the mixing-length parameter vary significantly, since there does not exist a physically sound reason to limit it \citep{Gough77}. Likewise, it would be poor practice to set the bounds of a uniform prior based on other numerical simulations. Thus we set the upper and lower bounds of $\alpha$ using the aforementioned empirical approach.

\begin{table}
\begin{center}
\caption{Lower and upper bounds used for the prior uniform densities for each stellar parameter.}
\label{tab:priors}
\begin{tabular}{@{}lcc@{}}
\hline
\hline
Parameter&  Lower bound & Upper bound\\
\hline
$M$ (M$_{\odot}$)      & 0.7& 1.25\\
$\stage$ (Gyr)        &  1& 13\\
$Z_0$                 & 0.010&  0.027\\
$X_0$                 & 0.525& 0.750\\
$\alpha$              & 1.0 & 3.5 \\
\hline
\end{tabular}
\end{center}
\end{table}

We finally assume that, besides $X_0$ and $Z_0$, all the parameters in the priors are uncorrelated, that is
\begin{equation}
\pi(\thetav) = \pi(M)\pi(\stage)\pi(\alpha)\pi(Z_0)\pi(X_0)\pi(\alpha_{\text{ov}}).
\end{equation} 

\subsubsection{Sampling method}\label{sect:MCMC}

Recovering the posterior density function $\pi(\thetav|\Xv)$ is the main technical issue of the estimation process. There is no closed-form solution to this inverse problem that gives $\thetav$ as a function of $\Xv$. We note that our use of the term {\lq}inverse problem{\lq} differs here from the more restricted scope encountered in solar physics \citep[see e.g.][]{JCD02}. We consider a much broader meaning as can be found for instance in \citet{Tarantola04} or \citet{Gregory05}.

Numerical methods are thus in order and in this paper we adopted an MCMC algorithm to carry out the estimation. The details of the method are given in Appendix~\ref{app:algo}. Their viability in the framework of stellar modelling has been discussed by \citet{Bazot08,Bazot12,Bazot16}. It has already been emphasised in these studies that fine tuning of an MCMC algorithm for stellar parameter estimation might turn out to be a subtle matter. These papers dealt with $\alpha$~Cen~A, which is a component of a binary system. As such, we have a strong prior on its mass, which greatly facilitates the sampling, restraining significantly the relevant space of parameters.

\begin{table*}
\center
\caption{Estimates of the stellar parameters of 18~Sco for $\Xv = (\teff, L, [\mathrm{Fe/H}], R, \{\delta\nu\}_{n,l})$. The first column gives the prior we used. Columns 2 to 6 give the mass, age, initial hydrogen mass fraction and metallicity and mixing-length parameter. The estimates for all but the age are the Maximum A Posteriori (see text) with the associated 68\% credible interval. The age was modelled with a Gaussian Mixture model and we give the mean, standard deviation and weight of each component. }
\label{tab:params-d02}
\centering                                      
\begin{tabular}{lccccccc}          
\hline\hline
\\[-3.mm]
Mass prior &$M/M_{\odot}$ & \multicolumn{3}{c}{$\stage$ (Gyr)} & $X_0$ & $Z_0$ & $\alpha$ \\
\hline
\\[-3.mm]
& & Mean & $\sigma$ & Weight & & \\
\cline{3-5}\\[-3.mm]
\multirow{4}{*}{Uniform}   &\multirow{4}{*}{$1.03_{-0.03}^{+0.03}$} &3.05 & 0.86 & 0.18 & \multirow{4}{*}{$0.716_{-0.024}^{+0.020}$} &\multirow{4}{*}{$0.0220_{-0.0010}^{+0.0011}$} &\multirow{4}{*}{$2.29_{-0.26}^{+0.54}$} \\
& & 4.75 & 0.69 & 0.36 & & & \\
& & 6.92 & 0.61 & 0.30 & & & \\
& & 8.87 & 0.88 & 0.16 & & & \\
\hline
\end{tabular}
\end{table*}

\subsection{Local optimisation}
\label{sect:orlagh}
It is important to compare the resulting parameters and their uncertainties using classical local optimisation methods with a Bayesian one.  The former are used widely in the literature and the shortcomings of such approaches need to be quantified and understood, and in particular, the (under-)determination of proper uncertainties.

Using the exact same code set-up as described in Sect.~\ref{sect:ASTEC}, we also used the Powell algorithm to find local solutions.  This algorithm has the advantage that it can minimise any function where the uncertainties on the data are not necessarily described by simple Gaussian distributions. Using the baseline dataset $\Xv = (\teff, L, [\mathrm{Fe/H}], R, \{\delta\nu\}_{n,l})$  (Sect.~\ref{sect:dataandphysicalmodel}) we proceeded to find local solutions by optimising the likelihood that appears in Eq.~(\ref{eq:bayes}).  The one difference between the two parameter sets $\theta$ is the use of the initial helium abundance $Y_0$ instead of the initial hydrogen abundance $X_0$ in the optimisation.  As $X_0 + Y_0 + Z_0 = 1$ this has no influence on the result.

Unlike the Bayesian approach, a local method suffers badly from correlations in parameters, that is it will find a solution close to the initial parameters if two of the free parameters are degenerate, whereas the Bayesian method will correctly extract all of these parameter correlations and additionally provide a better framework for interpreting the results.

Due to this local problem, we chose to work in a reduced 3-d parameter space to optimise $M, t_0, Z_0$ while fixing $X_0, \alpha$. Such an approach is typical and necessary \citep{Miglio05,Creevey07,Stello09,Creevey12,Dogan13,Lebreton14}. The optimisations were repeated using many initial guesses of the parameters as well as using different combinations of the latter. 

The estimation of our uncertainties is based on 1. generating a small grid around the optimal parameters, 2. perturbing the observations by their uncertainties 
and 3. finding the model from the grid that matches best to the perturbed observations.  These simulations were repeated 10,000 times, and we used the resulting distributions of 1-D parameters to describe their mean parameter and symmetric uncertainty.

\section{Results \& Discussion}\label{sect:blabla}
\begin{figure}[t!]
\center
\includegraphics[width=0.50\textwidth]{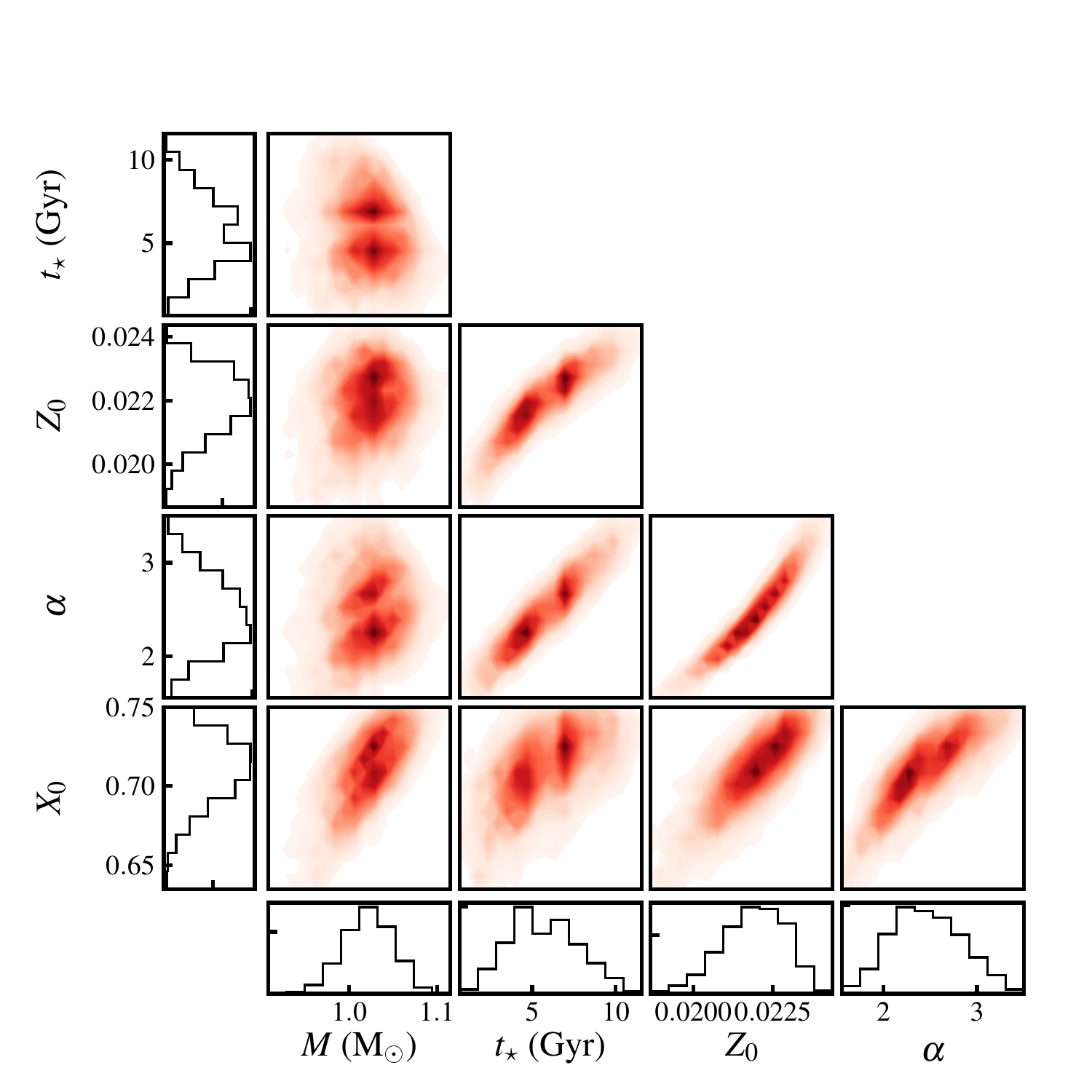}
\caption{Two- and one-dimensional marginal densities for the stellar parameters of 18 Sco for $\Xv = (\teff, L, [\mathrm{Fe/H}], R, \{\delta\nu\}_{n,l})$, with the uniform prior on the mass given in Sect.~\ref{sect:bsm}. The red shades in the central panels represent the two-dimensional marginal densities (normalised). The side panels represent the one-dimensional ones.}
\label{fig:2d-d02-mpru}
\end{figure}

\subsection{MCMC results}\label{sect:results}

In Appendix~\ref{app:algo}, we show that our MCMC simulations have converged to an acceptable level. We can thus merge the results from the independent chains we ran and obtain a posterior density for $\thetav$. The baseline case we analyse is for $\Xv = (\teff, L, [\mathrm{Fe/H}], R, \{\delta\nu\}_{n,l})$ and the uniform prior on the mass as explained in Sect.~\ref{sect:bsm}. In Table~\ref{tab:params-d02} we show the estimates for the individual stellar parameters based on the corresponding marginal densities. We also display the 68.3\% credible intervals intervals on the parameters. In the following, whenever a density is Gaussian the 68.3\% credible interval will be given using the symbol {\lq}$\pm${\rq}. In this case, the 68.3\% credible interval then coincides with the $1\sigma$ credible interval. Otherwise, the intervals are summed up as asymmetric error bars. These are defined as the smallest intervals containing the Maximum A Posteriori (MAP) and for which the posterior density integrates to 0.683 \citep[see][ and references therein]{Bazot16}.

It can be noted immediately that the mass estimate is in very good agreement with the one given in \citet{Bazot11}. This is particularly interesting since, as already noted, these have not been obtained using the same assumptions for the stellar model. Of course, without a proper modelling of the star, one cannot make statements on the other parameters. Nevertheless, it is a nice  \emph{a posteriori} confirmation for homology techniques \citep{Gough90}.

The two- and one-dimensional densities are shown in Fig.~\ref{fig:2d-d02-mpru}. Another notable feature in Fig.~\ref{fig:2d-d02-mpru} is seen in the side panels representing the one-dimensional marginal densities. These appear to be far from Gaussian. In particular the one for the age is multimodal. Consequently, establishing the statistical summaries such as those given in Table~\ref{tab:params-d02} demands some care. For unimodal distributions, we report the MAP estimate. The uncertainties are given as the smallest interval containing the MAP estimates for which the parameters have a 68.3\% probability to lie in. We define the level of precision as the estimated as the ratio of the length of the 68.3\% credible interval to the MAP estimate. This is a global estimate that does not account for any asymmetry in the density. For our baseline case, we obtained levels of precision of the order 6\% on the mass, 6\% and 9\% on $X_0$ and $Z_0$, and 35\% on $\alpha$. The precision on the age is discussed below in greater details.

\begin{figure}[t!]
\center
\includegraphics[width=0.50\textwidth]{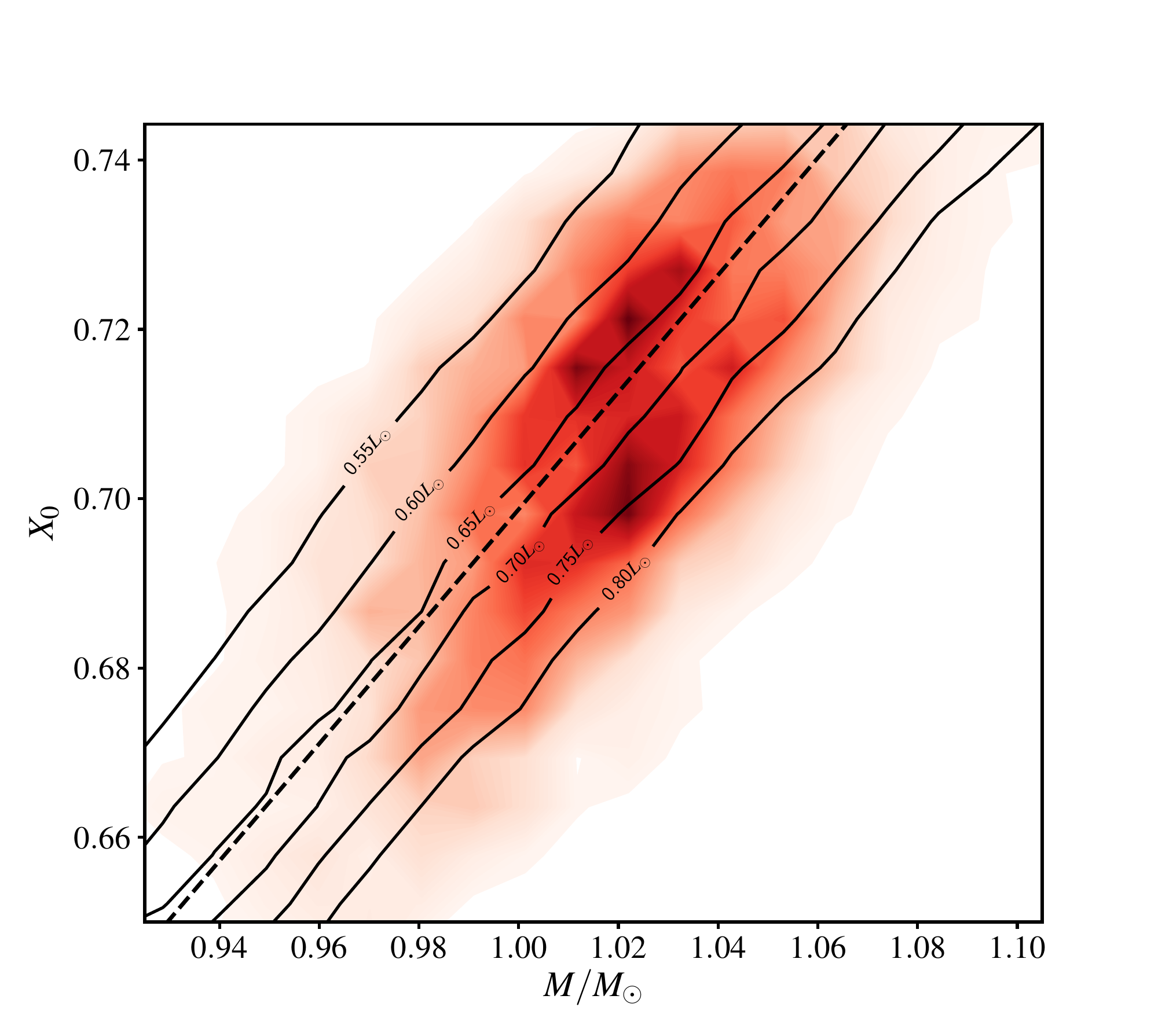}
\caption{Marginal joint probability density for $(M,X_0)$ (magnified from Fig.~\ref{fig:2d-d02-mpru}). The full lines mark locii of constant ZAMS luminosity. The dashed line shows the slope as obtained by Pearson's correlation coefficient.}
\label{fig:mxl}
\end{figure}

Examining the correlation coefficients for each pair of parameters shows that they all correlate to some degree. However, we can identify two groups that display significant correlations. As a rule of thumb, we consider as significant a correlation for which the Pearson's correlation coefficient \citep{Pearson1895} is $\gtrsim$ 0.7. First, the mass correlates strongly with $X_0$. Second, $\stage$, $\alpha$, $X_0$ and $Z_0$ are all tightly correlated. The correlation between the mass and $X_0$ can be explained by the adverse effects these two parameters have on the luminosity on the zero-age main sequence, $\lzams$. Mass-luminosity relations obtained from simple homology considerations \citep[see e.g.][]{Clayton68} establish clearly that an increase of $M$ induces an increase of $\lzams$. Conversely, the dependence of opacity on $X_0$, often assumed being a power law, implies that an increase in hydrogen-mass fraction corresponds to a luminosity decrease. In Figure~\ref{fig:mxl}, we display the posterior joint density for the couple $(M,X_0)$. Overplotted are lines of constant luminosity on the ZAMS and the slope obtained using Pearson's correlation coefficient. For this latter we retain its classical interpretation as the geometric average of the two regression slopes of $M$ by $X_0$ and $X_0$ by $M$ \citep{Rodgers88}. We see that all of these have the same direction. For that reason, we can associate the correlation between these two variables to be caused solely by the need to balance their effects on the initial luminosity. Since the mass does not correlate strongly with the other parameters, we can assume that once $\lzams$ is determined through the values of $Z_0$, $\alpha$ and $X_0$, then a value of $M$ is imposed.


The correlation between $X_0$ and $Z_0$ is intuitive and corresponds to setting the initial metal-to-hydrogen ratio so that the evolved star can reproduce the observed value of [Fe/H]. On a side note, we can notice that the behaviour of these two parameters is opposite to what is seen in \citet{Bazot16} in the case of $\alpha$~Cen~A. This is due to the fact that there exists a strong prior on the mass for this latter star. Therefore, it is not possible to set an adequate initial luminosity by varying $M$ and $X_0$ and this is instead achieved through relative adjustments of $X_0$ and $Z_0$. This shows how complex can be the dependence of the final posterior estimates on the precise functional form of the statistical model.

\begin{figure*}[t!]
\center
\includegraphics[width=0.475\textwidth]{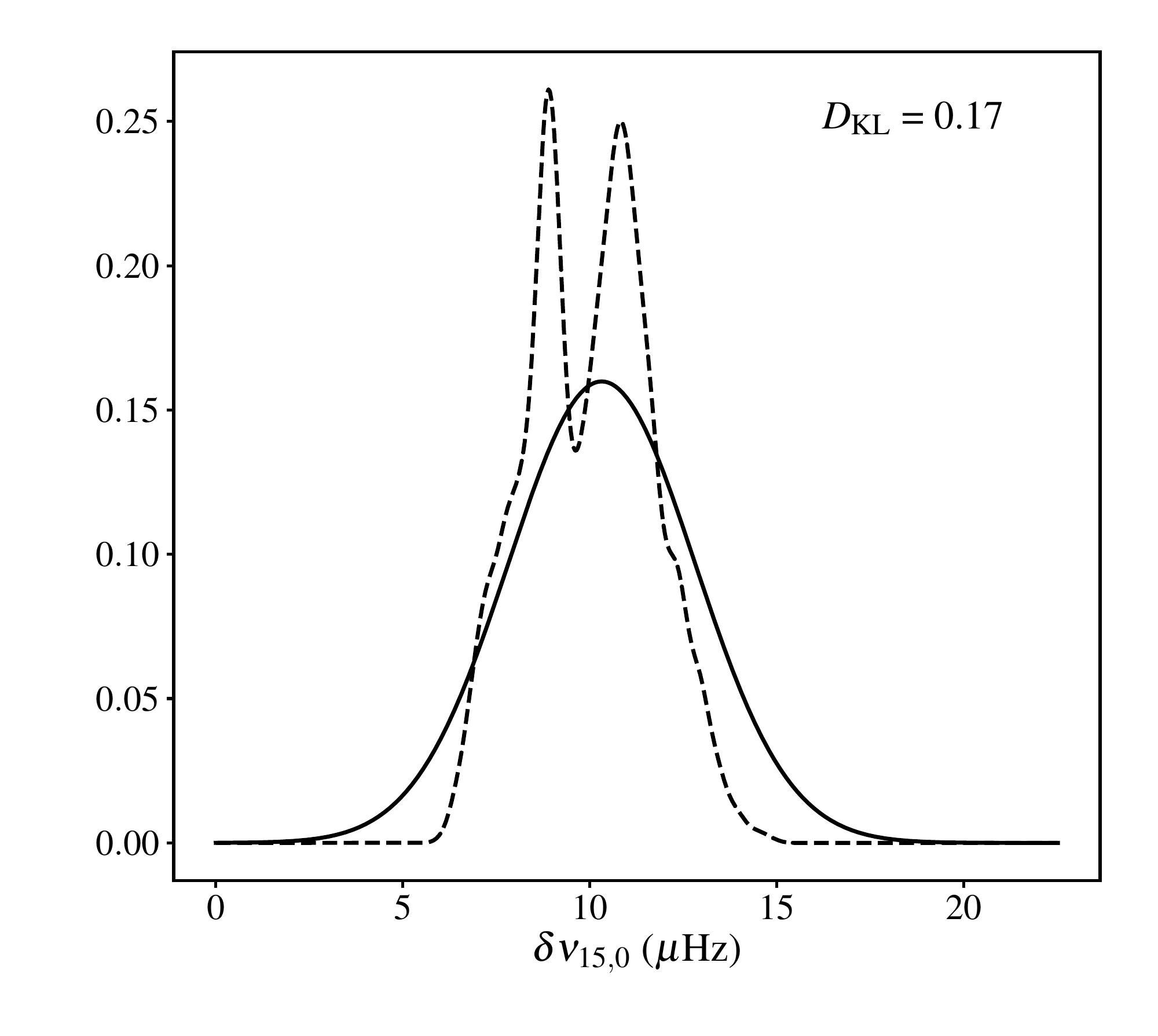}
\includegraphics[width=0.475\textwidth]{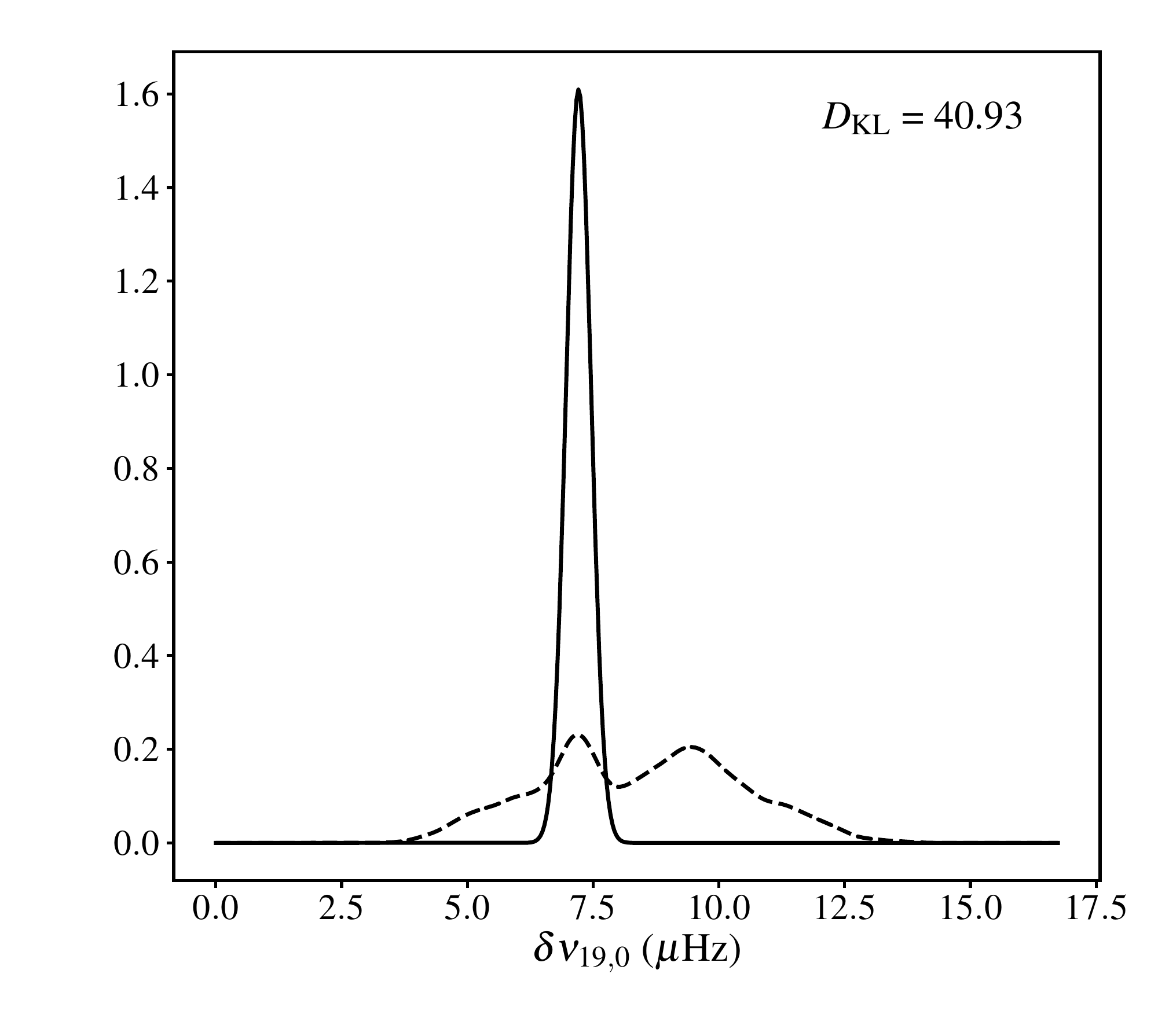}
\caption{Examples of observed (full lines) and posterior (dashed lines) probability densities for the small separations $\delta\nu_{15,0}$ (left panel) and $\delta\nu_{19,0}$ (right panel).}
\label{fig:ss-examples}
\end{figure*}

The correlations of $\alpha$ with $Z_0$ and $X_0$ can be partly explained by considering the effects of varying the mixing-length parameter. This affects mostly two characteristics of a stellar model: the depth of its convective zone and its effective temperature (and consequently its radius). The correlation between $\alpha$ and $Z_0$ thus sets the initial metal density in the convective zone, that is the ratio of $Z_0$ to the size of the convective zone. This is an important quantity because, together with the initial $Z_0/X_0$ it characterises the amount of metals that diffusive process ought to deplete the external convective zone in order to reproduce the observations. The correlation between $\alpha$ and $X_0$ is partly governed by the initial effective temperature. Because in the framework of the mixing-length theory an increase of $\alpha$ decreases the temperature gradient, then, all other things remaining equal, it will increase $\teff$. Decreases of $\teff$ induced from increases of $X_0$ can be deduced from homology relations which predict that the effective temperature is proportional to some power of the molecular weight. However, the initial effective temperature is not as strongly linked to the $(\alpha,X_0)$ correlation as is the ZAMS luminosity to the $(M,X_0)$. This indicates that other parameters, such as $Z_0$, have an influence on the initial $\teff$. Such intricate interplay are difficult to disentangle.

The same can be said for the correlation between the stellar age and $\alpha$, $Z_0$ and $X_0$. The age of the star is mostly controlled by the need to reproduce the internal layers of 18 Sco and thus its seismic characteristics. In that sense, it is the tight constraints on the age that imply the aforementioned correlations for the other parameters, as shall be discussed below. On that point suffice to say that Fig.~\ref{fig:2d-d02-mpru} shows that the age of the star increases with $\alpha$, $Z_0$ and $X_0$. The two first correlations are diffusion effects. Indeed, reaching the required amount of metal takes longer when $Z_0$ increases and diffusion becomes slower when, everything otherwise equal, the depth of the convective envelope becomes larger. The correlation with $X_0$ is related to the energetics of the star since the ZAMS luminosity is on average lower and that more time is necessary to reach the observed luminosity. 

\begin{figure}[t!]
\center
\includegraphics[width=0.475\textwidth]{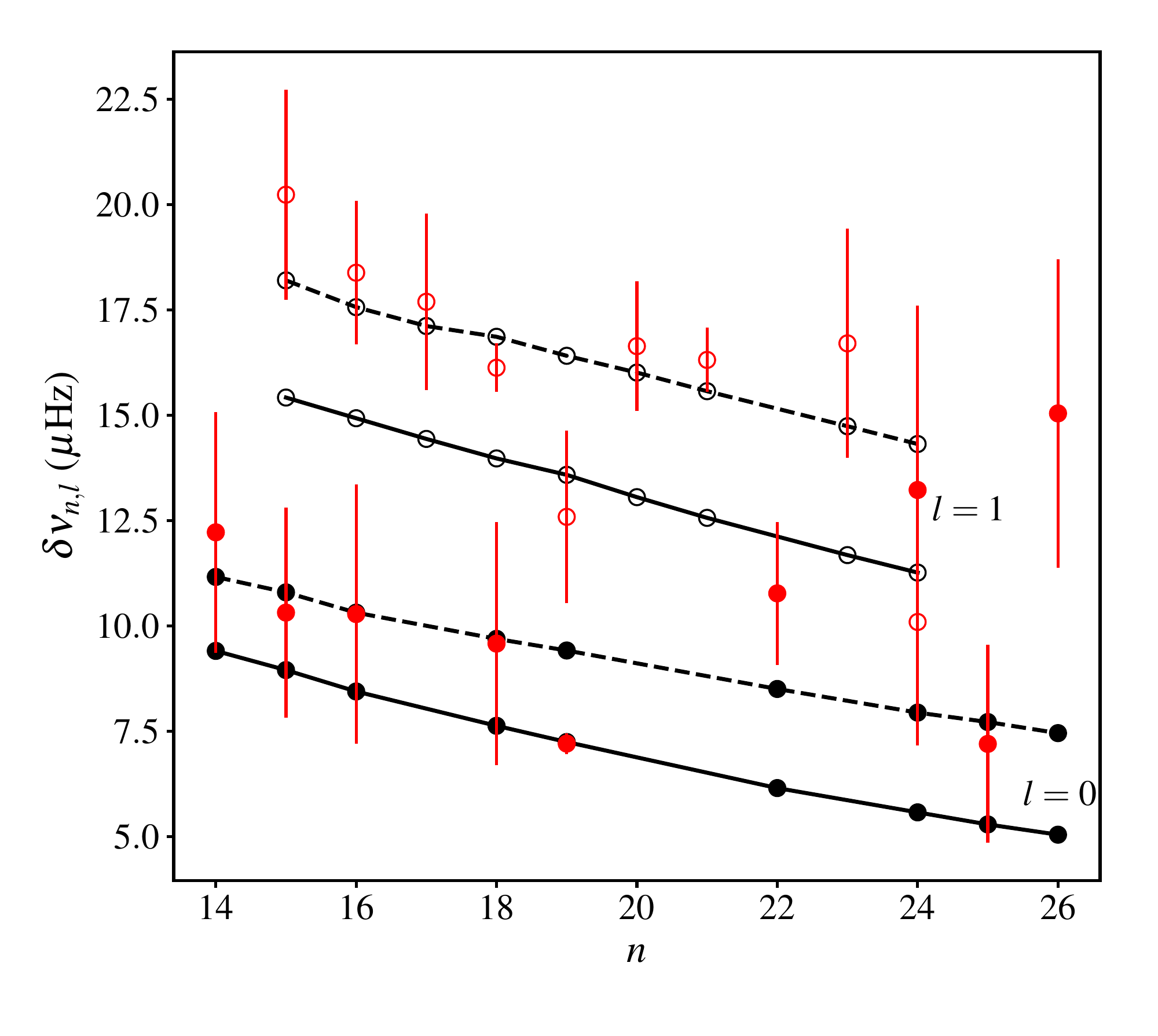}
\caption{Observed and estimated small separations. The dashed lines show the small separations of the best model with $\stage < 5$~Gyr. The full lines show the small separations of the best model with $\stage > 5$~Gyr. The red dots show the mean of the observed small separations and their standard deviations. The full dots mark small separations with $l=0$ and the open dots those with $l=1$.}
\label{fig:ss-modes}
\end{figure}

\begin{table*}
\center
\caption{Estimates of the stellar parameters of 18~Sco for different observational vectors $\Xv$. The format is similar to Table~\ref{tab:params-d02}}
\label{tab:params-others}
\centering                                      
\begin{tabular}{cccccccc}          
\hline\hline
\\[-3.mm]
$\Xv$ &$M/M_{\odot}$ & \multicolumn{3}{c}{$\stage$ (Gyr)} & $X_0$ & $Z_0$ & $\alpha$ \\
\hline
\\[-2.mm]
$\teff, L, [\mathrm{Fe/H}]$   &$1.01_{-0.10}^{+0.07}$ &\multicolumn{3}{c}{$5.64_{-3.33}^{+3.32}$} &$0.714_{-0.051}^{+0.032}$ &$0.0216_{-0.0015}^{+0.0011}$ &$2.09_{-0.29}^{+0.60}$ \\[1mm]
$\teff, L, [\mathrm{Fe/H}],R$ &$1.01_{-0.08}^{+0.06}$ &\multicolumn{3}{c}{$7.18_{-3.97}^{+2.89}$} &$0.713_{-0.048}^{+0.030}$ &$0.0222_{-0.0019}^{+0.0009}$ &$2.31_{-0.52}^{+0.47}$ \\[1mm]
\hline
\\[-2mm]
& & Mean & $\sigma$ & Weight & & \\
\cline{3-5}\\[-2.mm]
\multirow{4}{*}{$\teff, L, [\mathrm{Fe/H}],\{\delta\nu\}_{n,l}$} &\multirow{4}{*}{$1.02_{-0.09}^{+0.06}$} &2.57 & 0.87 & 0.21 & \multirow{4}{*}{$0.697_{-0.039}^{+0.045}$} &\multirow{4}{*}{$0.0213_{-0.0011}^{+0.0010}$} &\multirow{4}{*}{$2.17_{-0.35}^{+0.37}$} \\
& & 4.41 & 0.71 & 0.38 & & & \\
& & 6.59 & 0.74 & 0.28 & & & \\
& & 8.43 & 1.06 & 0.13 & & & \\
\hline
\end{tabular}
\end{table*}

The case of the age demands a more careful discussion. It has been established that its marginal density is bimodal. To account for this we used a Gaussian Mixture Model. We found that a reasonable agreement is obtained for four Gaussian components (this allows to accommodate reasonably well for the bimodality but also for clear asymmetries in the densities). One mode peaks at 4.46~Gyr and the other at 6.71~Gyr. We can describe both modes satisfactorily using two of the components given in Table~\ref{tab:params-d02}. The peak at 4.46~Gyr is well-described with the two modes with the smallest means and the one at 6.71~Gyr by the other two. Taking this into account, the weights of the lower- and upper-age modes are respectively 0.54 and 0.46. Therefore, one cannot conclude clearly on whether one of these two solutions is more likely than the other. Coming back to the problem of estimating credible intervals, one can separate the two peaks using the results of the Gaussian Mixture model fitting. In that case, it is necessary to renormalise the weights obtained, since only two components are used for each mode. Using such a procedure, we obtain as a MAP estimate $4.67_{-1.29}^{+0.87}$~Gyr. Likewise, for the upper-age mode, we derive a credible interval $6.95_{-0.89}^{+1.81}$~Gyr. The relative precisions are 46\% for the former mode and 39\% for the latter. We note that if we estimate the age in the sense of the Posterior Mean, then the two solutions are even further apart ($4.18\pm1.10$~Gyr and $7.60\pm1.17$~Gyr).

From what preceded, it seems obvious that the bimodality observed in the age marginal density of our main result stems from the very nature of the seismic data we used. If we look at the joint probability of the age and the theoretical individual small separations (not displayed here), we indeed see that they are strongly (anti-)correlated. In order to understand how the small-separation measurements affect the age, we need to examine the adequacy between the theoretical and observational density of the individual small separations. In short, we want to assess whether or not we could reproduce the seismic data.

Looking at these densities, one sees immediately that it is difficult to model them properly. Two examples are shown in Fig.~\ref{fig:ss-examples}. The MCMC-simulated densities always show a bimodality that maps the age bimodality. In order to get a feeling of the closeness between the observed and theoretical densities, we can compute the Kullback-Leibler distance, which is defined as the distance between two probability densities $p$ and $q$ \begin{equation}\displaystyle D_{KL}(p,q) = \int_{-\infty}^{+\infty}p(x)\log\left(\frac{p(x)}{q(x)}\right)dx. \end{equation}. The theoretical densities shown in Fig.~\ref{fig:ss-examples} are those with the smallest ($l=0,\ n=15$) and largest ($l=0,\ n=19$) $D_{KL}$. None of these reproduce perfectly the observed ones. However, for the case $l=0,\ n=15$ the absolute distance between the mean is 0.34 and the variance ratio is 2.08. Those values are 1.36 and 0.02 for the $l=0,\ n=19$ small separation. This indicates that the model reproduces much better the former than the latter.

It is in fact those small separations with the highest $D_{KL}$ that cause the bimodality of the age density. We sketch an explanation in Fig.~\ref{fig:ss-modes}. In there we plot the small separations obtained for the best models found in the MCMC sample for $\stage > 5$~Gyr (full line) and $\stage < 5$~Gyr (dashed line). For the sake of readability, we did not represent the full distribution of the small separations. Each corresponds to a different peak in the bimodal age density.  We also plot the observational means and variance to provide an idea of the agreement between these local best models and the observed densities of the small separations. We see there that only five observed small separations are compatible with both models. Of the 13 remaining small separations, two are far closer to the older model, in particular $\delta\nu_{19,0}$, which is likely to control the old-age solution. The other ones, are mostly compatible with the lower-age solution, even though they have larger variances than the distribution of $\delta\nu_{19,0}$ and thus only partially compensate the impact of this latter.
\begin{figure}[t!]
\center
\includegraphics[width=0.5\textwidth]{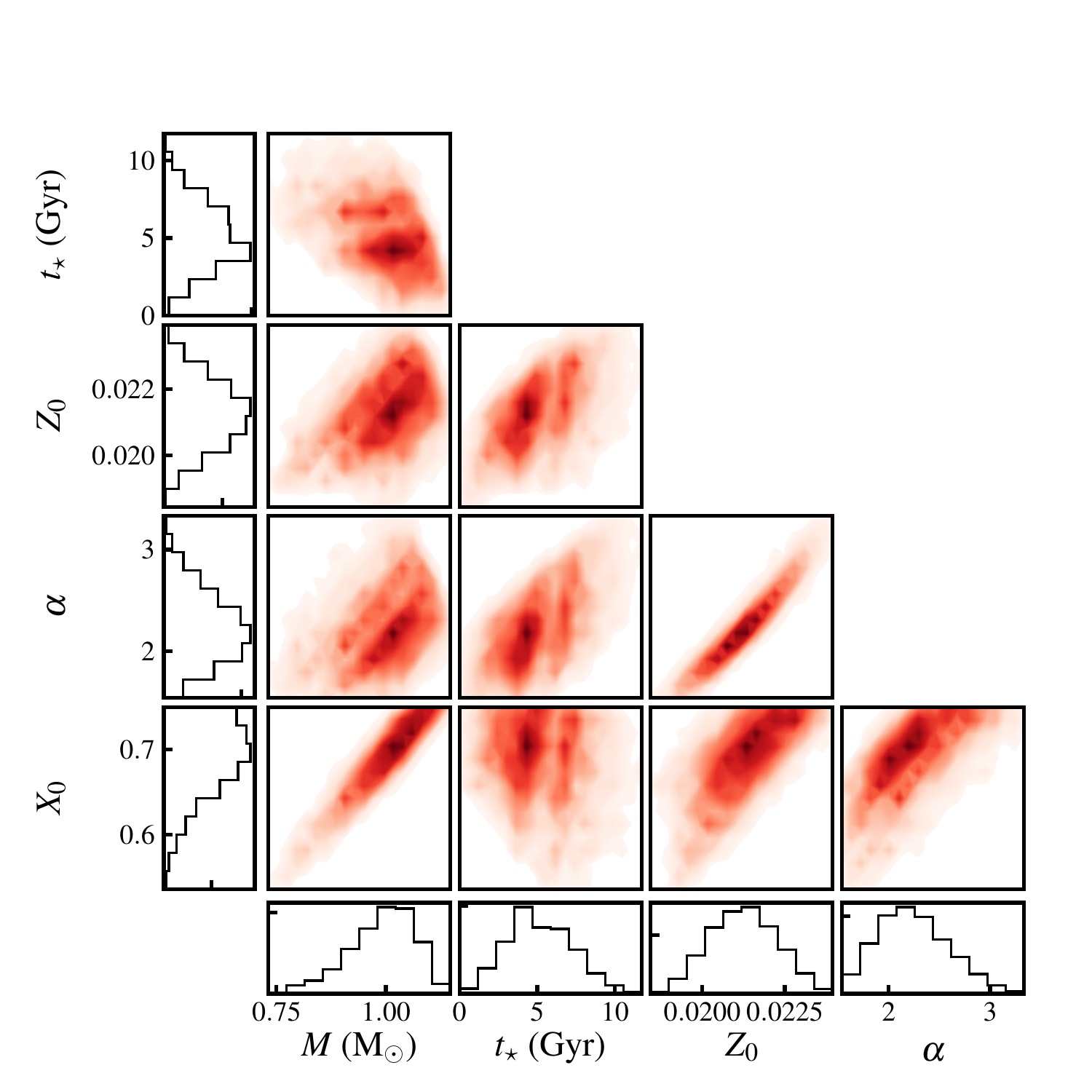}
\caption{Same as Fig.~\ref{fig:2d-d02-mpru} but with $\Xv  = (\teff, L, [\mathrm{Fe/H}],[\delta\nu]_{n,l})$. }
\label{fig:2d-nr-mpru}
\end{figure}

From this discussion, we can conclude that the main source of error in our result does not come from the modelling of the observational errors. Rather, it is the data themselves that impose limitations on our analysis, either because of their intrinsic properties, that is the noise, or because their modelling in \citet{Bazot12} was not accurate enough. The density of the small separations as we could derive them from MCMC samples in Sect.~\ref{sect:sismo} suggests larger variations with the mode order (or the frequency) than what the stellar models can accommodate. Therefore, using only the data at hand cannot allow us to make a choice on which solution for the stellar age is the preferred. Either new data or a re-analysis of the existing time series with more adequate techniques is needed to go further. 

\subsection{Effect of observational constraints}

\begin{figure*}[t!]
\center
\includegraphics[width=0.45\textwidth]{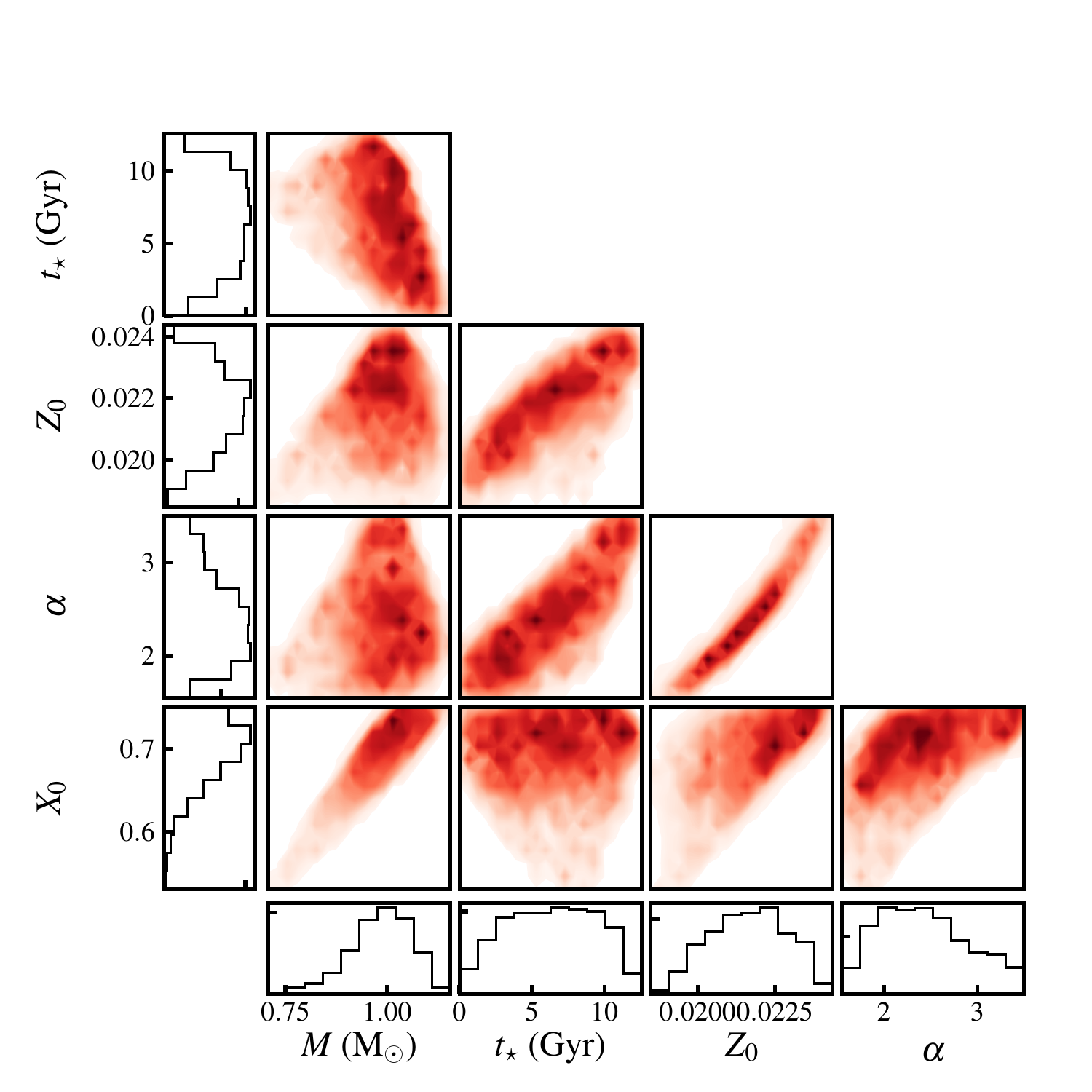}
\includegraphics[width=0.45\textwidth]{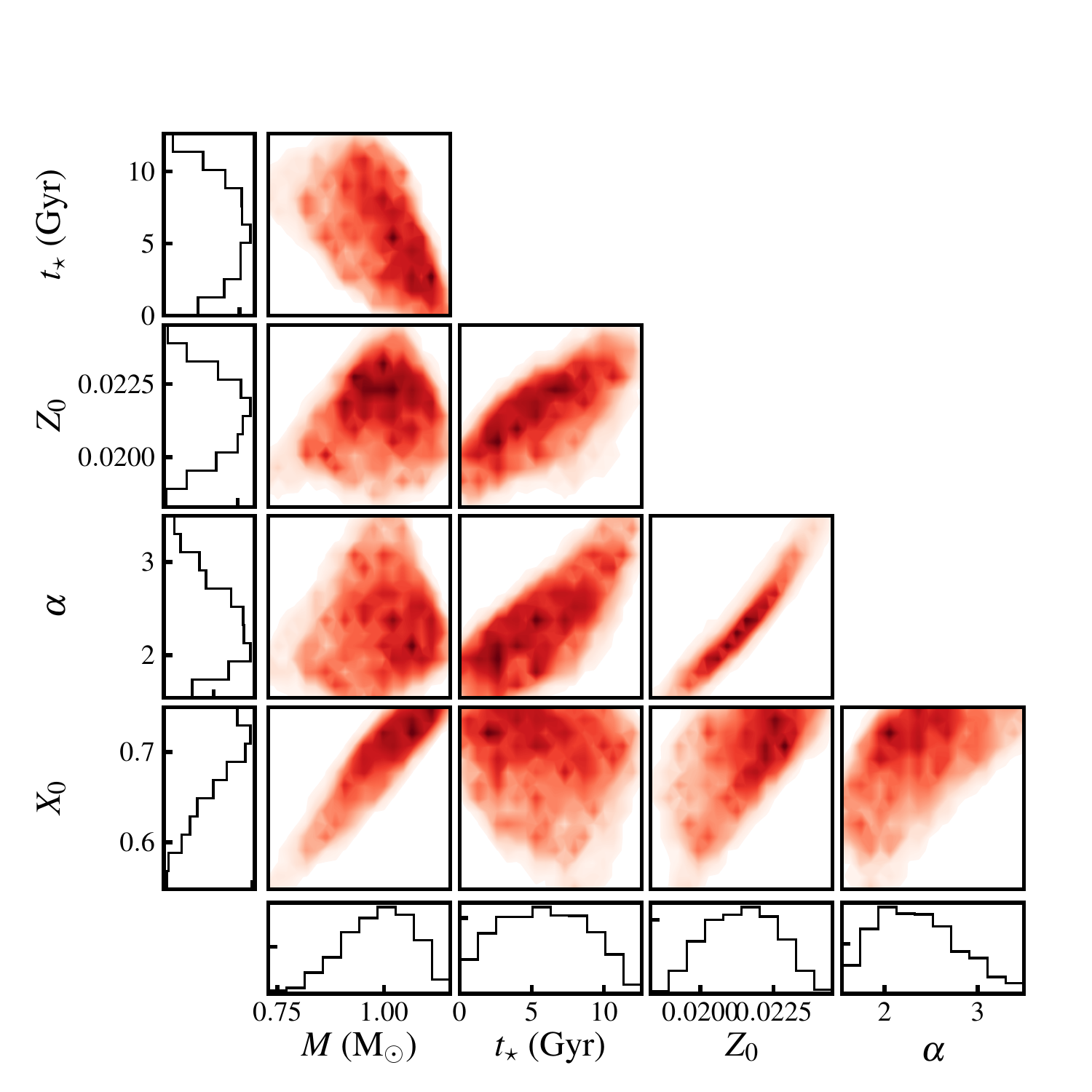}
\caption{Same as Fig.~\ref{fig:2d-d02-mpru} but with $\Xv  = (\teff, L, [\mathrm{Fe/H}],R)$ and $\Xv  = (\teff, L, [\mathrm{Fe/H}])$ in the left and right panel, respectively. }
\label{fig:2d-nsis-mpru}
\end{figure*}

In order to test the effect of the observations as constraints, we also ran three MCMC simulations with different observation vectors: $\Xv = (\teff, L, [\mathrm{Fe/H}], \{\delta\nu\}_{n,l})$, $\Xv = (\teff, L, [\mathrm{Fe/H}], R)$, $\Xv = (\teff, L, [\mathrm{Fe/H}])$. The resulting estimates of the parameters are given in Table~\ref{tab:params-others}. In Figure~\ref{fig:2d-nr-mpru} we show the two- and one-dimensional marginal densities of the stellar parameters for the case $\Xv = (\teff, L, [\mathrm{Fe/H}], \{\delta\nu\}_{n,l})$. In Figure~\ref{fig:2d-nsis-mpru} we display the same graphs for the two cases in which the seismic data were not included. We did not split up the spectro-photometric data since we considered that without these basic observations together, it is not possible to get any useful estimate of the stellar parameters. As discussed below, this assumption seems justified \emph{a posteriori}.

The case $\Xv = (\teff, L, [\mathrm{Fe/H}], \{\delta\nu\}_{n,l})$ shows the effect of adding a radius measurement to the basic spectro-photometric data. We first note that the bimodal structure of the age density is preserved. Second, we see a significant decrease of the precision on the mass, which now drops to 15\%. This a consequence of the loss of information on the average density of the star \citep{Creevey07,Bazot11}. Given the existing correlation between $M$ and $X_0$, it is unsurprising that precision also decreases for this latter parameter, down to 12\%. However, the decrease in precision is not as large as it is for the stellar mass. An explanation is that the lack of constraint on the final average stellar density allows for correlation between the mass and other stellar quantities besides $X_0$. We indeed see in Fig.~\ref{fig:2d-nr-mpru} that it now correlates marginally with $\stage$, $\alpha$ and $Z_0$. The correlation with the age is a well-known trend in stellar physics. On isochrones, luminosity and effective temperature increase with the stellar mass. Therefore, to reproduce a similar data set with a larger mass, one needs to decrease the age. In retrospect this behaviour sheds light on the very small correlation seen between $M$ and $\stage$ seen in Sect.~\ref{sect:results}. It indicates that, in the regime defined by the observations of 18~Sco, the age-mass relation is not as steep as the mass-$X_0$ relation and that the effects of the former can only be seen when $M$ is allowed to vary on wider ranges. The correlations with $\alpha$ and $Z_0$ reflects, as above, the need to set up the ZAMS model adequately, only this time with one more degree of freedom. Rather than being fixed by balancing $M$ and $X_0$, the initial luminosity is now the result of an interplay between $M$, $\alpha$, $X_0$ and $Z_0$. Consequently, we do not observe such a good alignment between lines of constant $\lzams$ and the geometric average of the regression lines in the $(M,X_0)$ plane as seen in Fig.~\ref{fig:mxl} for the case including the radius measurement.

The precisions on $\alpha$ and $Z_0$ do not change, even though the MAP estimates do. This stresses that the radius does not provide such an important constraint on these parameters. It was already noticed by \citet{Creevey07} that understanding the relation between the error on the radius and the uncertainty on $\alpha$ is difficult and depends on the details of the model. In the case of 18~Sco, this relative independence can be understood by the fact that the critical quantity upon which $\alpha$ and $Z_0$ act is not the radius, but the density of metals in the convective zone. This is not constrained by the radius but rather by the surface [Fe/H] ratio. Looking at the results for $\Xv = (\teff, L, [\mathrm{Fe/H}], R)$ and $\Xv = (\teff, L, [\mathrm{Fe/H}])$ in Fig.~\ref{fig:2d-nsis-mpru} seems to confirm this. First we see that the age distribution does not display any significant bimodality. The loss of precision on the age compared with cases that include seismic data is very significant. If the radius is included precision is of the order 96\%, otherwise it is of the order of 118\%. This implies a small loss of precision in $\alpha$, but not of the same magnitude. There is, however, almost no loss of precision in $Z_0$ and $X_0$. In Figure~\ref{fig:2d-nsis-mpru}, the two-dimensional marginal PDF that preserves a structure relatively similar to those seen in Figs.~\ref{fig:2d-d02-mpru} and \ref{fig:2d-nr-mpru} is the one for $(\alpha,Z_0)$. This sheds light on the the role of the very precise measurement of [Fe/H] that exists for 18~Sco. This true regardless of whether or not the radius has been included as a constraint.

\begin{figure*}
\center
\includegraphics[width=\textwidth]{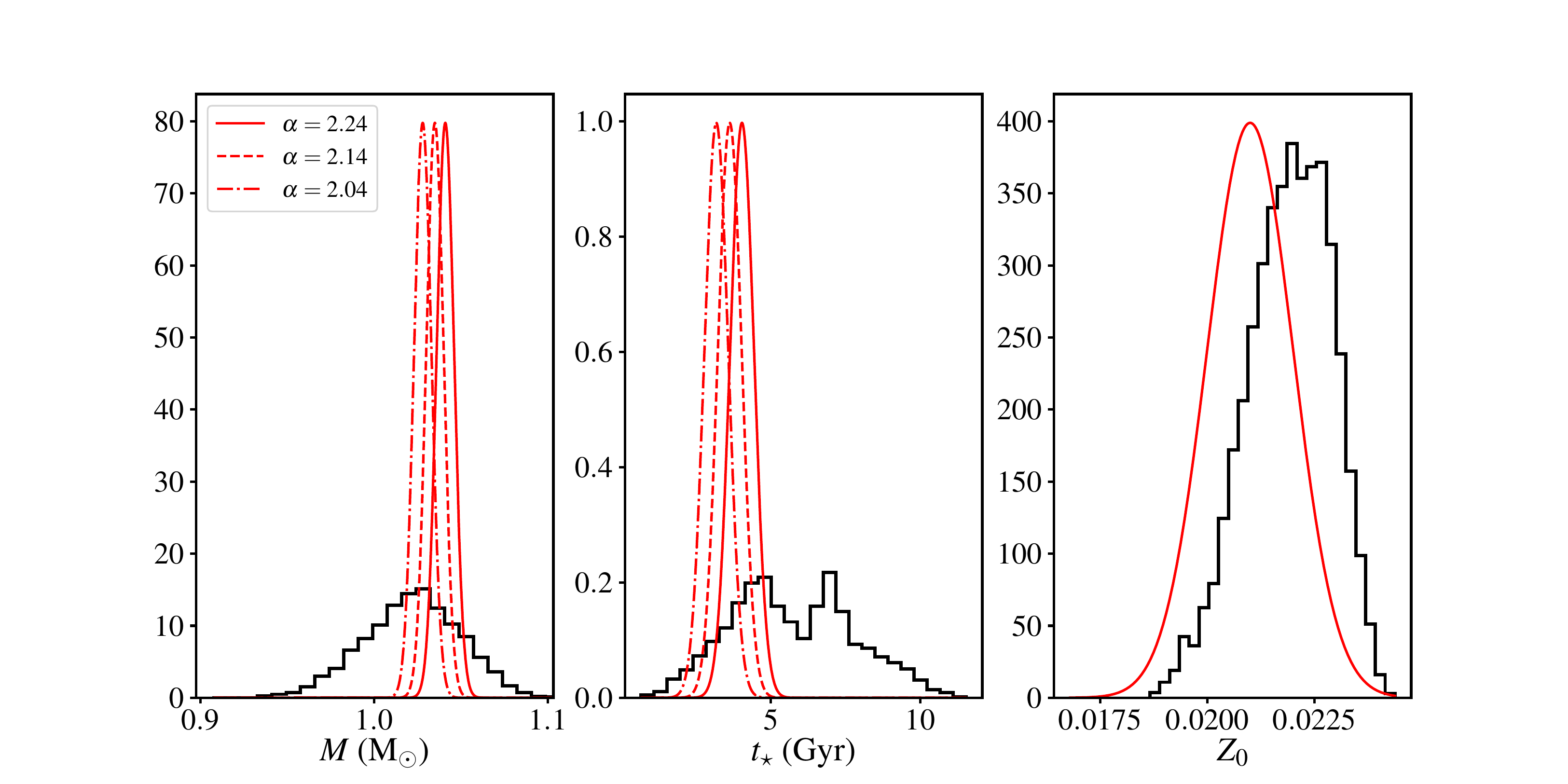}
\includegraphics[width=\textwidth]{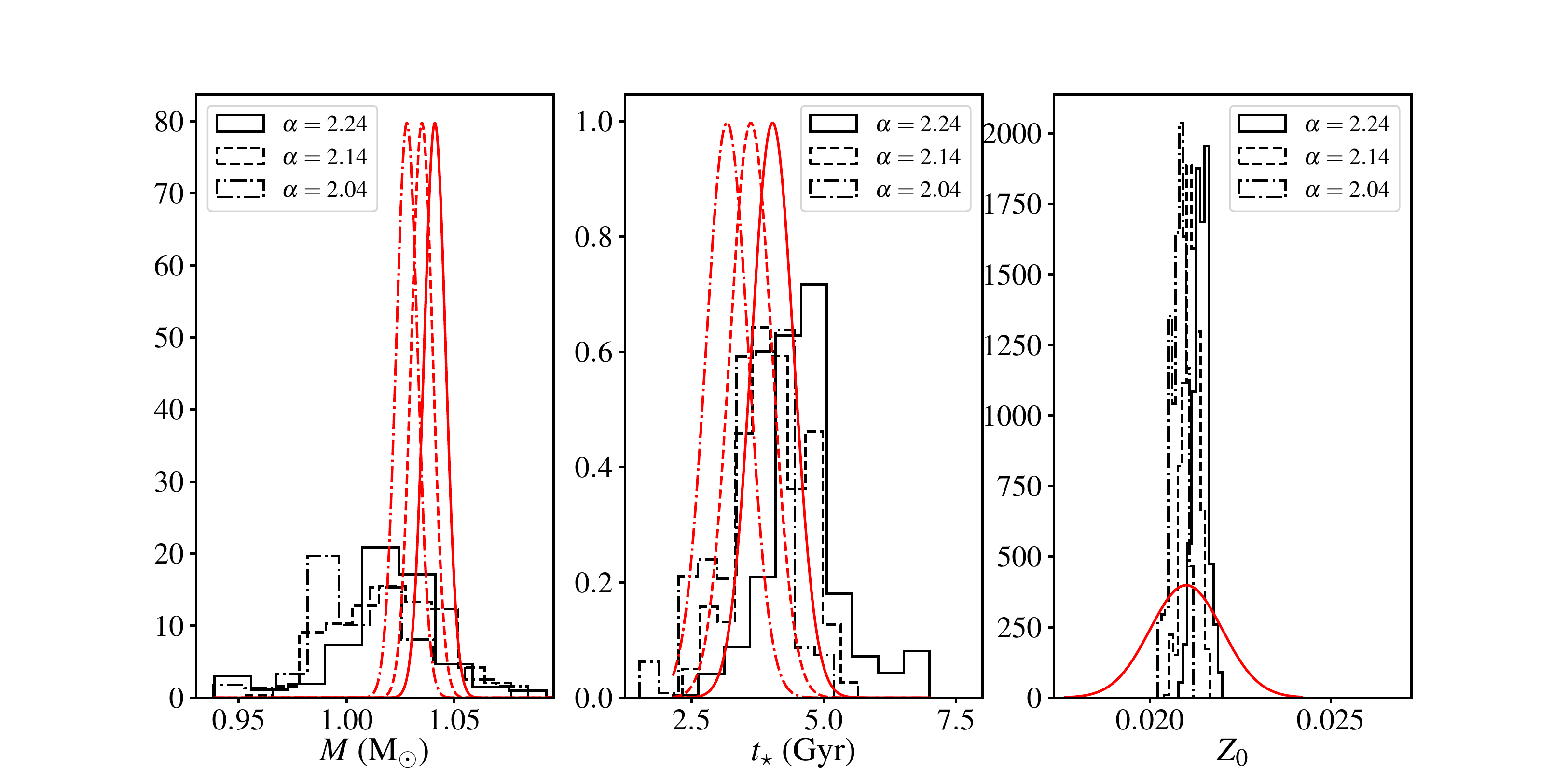}
\caption{Upper row: Marginal distributions (black) of $M$, $t$, and $Z_0$ from the results presented in Fig.~\ref{fig:2d-d02-mpru} (Sect.~\ref{sect:blabla}), and the distribution for these same parameters as determined from the local analysis (red Gaussians, Sect.~\ref{sect:orlagh}, Table~\ref{tab:localparameters}). Bottom row: Marginal distributions for the same parameters, but restricted to a small interval around the values of $\alpha$ given in Table~\ref{tab:localparameters}. The red lines are similar to those in the upper row.}
\label{fig:comparegloballocal}
\end{figure*}

\subsection{Comparison with local optimization}\label{sect:comparison}

\begin{table}
\center
\caption{Results from local optimisation for the reference case $\Xv = (\teff, L, Z/X, R, \{\nu_{n,l}\})$. The first three columns give the Maximum Likelihood Estimates for the $M$, $\stage$ and $Z_0$. The last two columns give the values to which $X_0$ and $\alpha$ were fixed to perform the optimisation. The last row gives the estimated uncertainties, expressed as the standard deviation of a Gaussian distribution.}
\label{tab:localparameters}
\centering
\begin{tabular}{ccccc}
\hline\hline
\\[-3.mm]
$M/M_{\odot}$ & $\stage$ (Gyr) & $Z_0$ & $X_0$ & $\alpha$ \\
\hline
\\[-3.mm]
 $1.041$  &$4.09$    &$0.021$ & $0.711$ & $2.24$ \\
 $1.035$  &$3.65$    &$0.021$ & $0.702$ & $2.14$ \\
 $1.028$  &$3.22$    &$0.021$ & $0.692$ & $2.04$ \\
\hline
 $0.005$  &$0.40$    &$0.001$ & -- & -- \\
\hline
\end{tabular}
\end{table}

After presenting the results from the Bayesian approach, we aim to compare the resulting model parameters with those obtained from local optimisation and attempt to quantify the differences in the uncertainties using the two approaches. Optimal models and uncertainties from the local method are given in Table~\ref{tab:localparameters}. These have been obtained for the baseline case with $\Xv = (\teff, L, Z/X, R, \{\nu_{n,l}\})$. As can be noted the uncertainties are much smaller compared to those presented in Table~\ref{tab:params-d02}, being based on fixing $X_0$ and $\alpha$. In all cases, the optimal parameters are in good agreement with the marginal distributions obtained from the MCMC simulations. Comparing these values directly with Fig.~\ref{fig:2d-d02-mpru}, we can see that by restricting $0.692 < X_0 < 0.711$ ($0.287 >Y_0 > 0.268$) the local uncertainties that we obtain are not entirely underestimated. For $X_0 = 0.692$ for example, solutions are found between 1.00 and 1.05 M$_{\odot}$, in agreement with the values of $M = 1.041$ M$_{\odot}$, $1.035$ M$_{\odot}$ and $1.028$ M$_{\odot}$ proposed  in Table~\ref{tab:localparameters}.

It is interesting to visualise the real discrepancy between them knowing that our assumptions on the local analysis are indeed unrealistic. There is no possibility to constrain the values of $\alpha$ and $X_0$ given the current (and most likely future) observations used in this work. The upper row in Fig.~\ref{fig:comparegloballocal} illustrates the 1-D marginal distributions for $M$, $t_{\star}$ and $Z_0$ as inferred from the Bayesian analysis (black curves), along with the inferred parameters and uncertainties from local optimisation. The parameters and uncertainties from the local analysis are represented by the red Gaussian distributions for the three solutions provided in Table~\ref{tab:localparameters}. We note that in particular, we entirely fail to obtain a solution in the second age range proposed by the Bayesian approach.

Some of this discrepancy can be explained by the need to fix some parameters in the local optimisation procedure. To test this we restricted the MCMC sample to ranges $\alpha\pm 0.01$ for all cases. We see in the bottom row of of Fig.~\ref{fig:comparegloballocal} the corresponding distributions for all three values of the mixing-length parameter given in Table~\ref{tab:localparameters}. A significant agreement is then reinstated for the age, although this means that the large-age solution observed in the original sample has been filtered out by our cut on $\alpha$. Likewise, the mass distribution is also much closer to the one estimated from local optimisation. However, the metallicity density, whose estimates obtained from the local optimisation and MCMC strategies looked fairly similar, now becomes much narrower in the former case.

This highlights the difficulty there is to find a proper agreement between the two approaches. It is extremely difficult to rule out any similarity between the two outcomes as not being incidental. This is due to the fact that, while the dimension of the problem increases, it becomes more difficult to keep track of correlations between parameters This is the so-called curse of dimensionality.  To that issue, MCMC algorithms offer a better operational solution, due to their ability to explore stochastically the space of parameters. They could potentially be used to serve as a benchmark for less-time-consuming local optimisation strategies.

Of course, this explanation of the discrepancy in terms of fixed parameters does not account for the difficulty to identify a second mode in the marginal age PDF using local optimisation. In this case, optimisation algorithms, which provide, by definition, point estimates, naturally underperform. 

\subsection{Other stellar parameter estimates}

In Sect.~\ref{sect:comparison} we have estimated the stellar parameters using a local optimisation strategy. The estimates there are in agreement with all those presented in the previous section. However, the relevant quantities are not only the point estimates for the parameters but also the uncertainties one can associate to these values. Those quoted in Tables~\ref{tab:localparameters}, \ref{tab:params-d02} and \ref{tab:params-others} are consistent for $Z_0$. For the other parameters, they differ much, sometimes close to an order of magnitude. It is noteworthy that the uncertainties obtained from optimisation do not vary much when the observation vector $\Xv$ changes. This indicates that a lot of information is factored in the assumptions made in Sect.~\ref{sect:orlagh} for the derivation of the uncertainties and that this may lead to underestimating them.

We have seen above that it is hard to reproduce perfectly the seismic data, and that its inclusion leads to a double solution to the estimation problem. One could extrapolate to a case in which more accurate seismic data would be available and expect lower uncertainties on the age, that could come closer to those of Table~\ref{tab:localparameters}. Nevertheless, even in that case, a proper modelling of the uncertainties points towards uncertainties of the order of $\sim$1~Gyr, which remains more than twice those found with optimisation. The convergence results from Appendix~\ref{app:algo} show that the samples generated from the MCMC are reliable, therefore one should clearly study carefully the details of the estimation strategy chosen to obtain stellar parameters before trusting the uncertainties. Contrary to what has been done in previous so-called {\lq}hare-and-hounds{\rq} exercises, this comparison only focuses on the differences in the methodology used to obtain the parameters. This means that we have used the exact same data and code (and code setup), which is not always the case in other comparison studies \citep[see e.g.][]{Reese16}.

Comparison with previously derived stellar parameters for 18~Sco is difficult precisely for this reason. It becomes very hard to disentangle the effect of the estimation strategy, the data that constrain the model and the precise numerics of the stellar evolution code used. We can point out a few recent estimates given in the literature, limiting ourselves to the age, which is the parameter the most difficult to assess but that could be crucial, in particular for studies that focus on Li depletion on the main sequence \citep{Israelian09,Melendez10}. \citet{Carlos16} give an estimate of $3.8\pm0.5$~Gyr. This was obtained by comparison with the Yonsei-Yale \citep{Kim02} set of isochrones  and using only spectrophotometric constraints. A notable difference is that $\log g$ was considered instead of the luminosity. The estimates of the atmospheric parameters and their associated uncertainties also differ slightly. However, it remains extremely unlikely that these changes could account for the difference with the uncertainties in the range 3 -- 4~Gyr obtained with $\Xv = (\teff, L, [\mathrm{Fe/H}])$. At any rate, this should not allow to obtain uncertainties lower than those obtained using seismic data. Other recent estimates can be found in \citet{Ramirez14} and \citet{Spina18}, they give $\stage = 3.0^{+0.3}_{-0.6}$~Gyr and $\stage = 4.2^{+0.3}_{-0.5}$~Gyr respectively. These uncertainties are again much lower, by an order of magnitude, than what is found using our method. These results were also obtained using isochrone fitting procedures, together with a Bayesian Statistical model. Only spectrophotometric parameters were considered, \citet{Spina18} using both $\log g$ and the luminosity. A likely explanation for such a discrepancy is the difficulty to sample properly the space of stellar parameters using pre-computed isochrones \citep{Bazot12}. Therefore, some models are not taken into account either due to incomplete sampling or because some stellar parameters have been fixed, thus reducing the final variance. As a sanity check we notice that the models from the MCMC simulation reproduce satisfactorily the probability density of the atmospheric parameters, indicating that a wide range of stellar ages can indeed account for such a combination.

Other studies provide ages derived using stellar population statistics. Some have focused on the so-called chemical-clocking methods, which are based on the chemical evolution of our galaxy, that is how much the interstellar medium from which the star was formed was enriched in Y, Mg and Al. \citet{TucciMaia16} estimate the age to be $3.090\pm0.391$~Gyr based on  an average age-[Y/Mg] relationship. \citet{Spina18} found age estimate ranging from $3.2 \pm 0.9$~Gyr to $4.3 \pm0.5$~Gyr depending on whether they use an age-[Y/Mg] or an age-[Y/Al] relationship and on the precise nature of their fit. Interestingly, \citet{Nissen17} seem to confirm these relationships using the Kepler LEGACY database \citep{SA17}.

Finally, some other studies focused on age-activity relationship to provide an estimate to $\stage$ for 18~Sco. Noteworthy are \citet{Mittag16} and \citet{LO18} which give respectively $5.1\pm1.1$~Gyr and $4.6\pm0.9$~Gyr. Similarly to the chemical-clocking estimates, the point estimates for the age are compatible within their error bars. However, the critical point is that those error bars differ significantly from the ones found using direct modelling and the Bayesian Statistics approach coupled to MCMC sampling. This raises the question of understanding how average-based estimates are representative of single objects. In other words how are these estimates affected by systematics. One also needs to understand if the current estimates for the age provided in this paper hint at the need to recalibrate these relations using more realistic uncertainties.



\section{Conclusion}

In this paper, we obtained estimates of the physical parameters of the solar twin 18~Sco using existing seismic data. Special care was taken to describe the asteroseismic diagnostic for the stars. A Bayesian Statistical model was used to relate the observations and the stellar parameters and statistical samples were obtained using an MCMC algorithm. A bimodal solution is obtained for the age, due to the difficulty to reproduce the seismic data. The most likely result gives an age that is roughly solar. This also points out the limitation of the current ground-based seismic data. This result may thus be used has benchmarks to evaluate in a near future the improvements made using the forthcoming TESS data or, potentially, SONG measurements.

Comparison of the resulting uncertainties with those obtained from local optimisation shows a discrepancy, the MCMC simulations leading to much larger uncertainties. The same conclusion applies when comparing these results to previous estimates in the literature. This motivates a more thorough investigation of the strategies used to estimate uncertainties on the physical characteristics, and in particular the age, of other solar twins. 

\begin{acknowledgements}
MB would like to thank S. Hannestad for providing him access to the Grendel cluster at DCSC/AU of which important use has been made during this work. This material is based upon work supported by the NYU Abu Dhabi Institute under grant G1502. Part of this research was supported through the Laboratoire Lagrange BQR funding scheme. Funding for the Stellar Astrophysics Centre is provided by The Danish National Research Foundation (Grant DNRF106).
\end{acknowledgements}

\bibliography{18scoref}

\appendix
\begin{algorithm}
  \caption{Multiple SA-AMGAS algorithm.}
  \label{alg:SA-AMGAS}
  \begin{algorithmic}
    \State $m = 1$
    \While {$m \leq M$}
    \State Generate $\thetav^{(0)}_m \sim \pi(.)$
    \State $t = T$
    \While {$t \geq 0$}
    \State Compute $\pi(\thetav^{(0)}_m)\pi(\Xv|\thetav^{(0)}_m)^{1/2^t}$
    \State $n = 1$
    \While {$1 \leq N$}
    \State Generate $\thetav^{\ast}_m \sim q(.|\theta^{(n-1)}_m)$
    \State Generate $\rho \sim \mathcal{U}([0,1])$
    \If {$\rho \leq \max\left(\frac{\pi(\thetav^{\ast}_m)\pi(\Xv|\thetav^{\ast}_m)^{1/2^t}}{\pi(\thetav^{(n-1)}_m)\pi(\Xv|\thetav^{(n-1)}_m)^{1/2^t}} , 1\right)$}
    \State Set $\thetav^{(n)}_m = \thetav^{\ast}_m$
    \Else
    \State Set $\thetav^{(n)}_m = \thetav^{(n-1)}_m$
    \EndIf
    \State n = n + 1
    \EndWhile
    \State $\thetav^{(0)}_m = \thetav^{(N)}_m$
    \State t = t - 1
    \EndWhile
    \State m = m + 1
    \EndWhile
\end{algorithmic}
\end{algorithm}

\section{MCMC sampling}

\subsection{Algorithm}\label{app:algo}

For the sake of completeness we describe here the MCMC algorithm used in this work. In order to explore efficiently the space of parameters, we combined a Simulated Annealing \citep[SA,][]{Liang10} algorithm with an Adaptive Markov chain Monte Carlo algorithm with Global Adaptive Scaling \citep[AMGAS,][Algorithm 4]{Andrieu08}. The pseudo-code is given in Algorithm~\ref{alg:SA-AMGAS}. In there, the quantity $q(.|\theta^{(n-1)}_m)$ is the proposal density used by the MCMC algorithm to obtain a trial parameter, conditional on the current value of the Markov chain. Additionally, we ran several Markov chains in parallel. 

The SA and AMGAS components of the algorithms are expected to improve the efficiency of the sampler. The former will improve convergence to the regions of highest probability density, even when the initial guess is chosen close to another local minimum. The latter improves the efficiency of the classical Metropolis-Hasting algorithm \citep{Metropolis53,Hastings70} by improving iteratively the proposal density of the algorithm. The adaption of the scaling factor is related to the idea of of optimal scaling \citep{Rosenthal08} and allows to adapt the scaling factor of the (Gaussian) proposal density so that it approaches a certain acceptance rate.

The main idea behind running multiple chains is to improve convergence diagnostics of MCMC algorithm \citep{Gelman92,Brooks98}. In principle, multiple chains do not sample the space of parameters better than one single long chain. In practice though, in the special case of stellar models, which take a long time to compute, it is beneficial to construct a sample from different chains, provided we can assess with some confidence that these have converged, since we are able to share the computational load on several processors. This helps the post-processing analysis of the posterior densities. In particular, the modelling of densities is improved when the size of the sample increases.

\begin{figure}[t!]
\center
\includegraphics[width=0.475\textwidth]{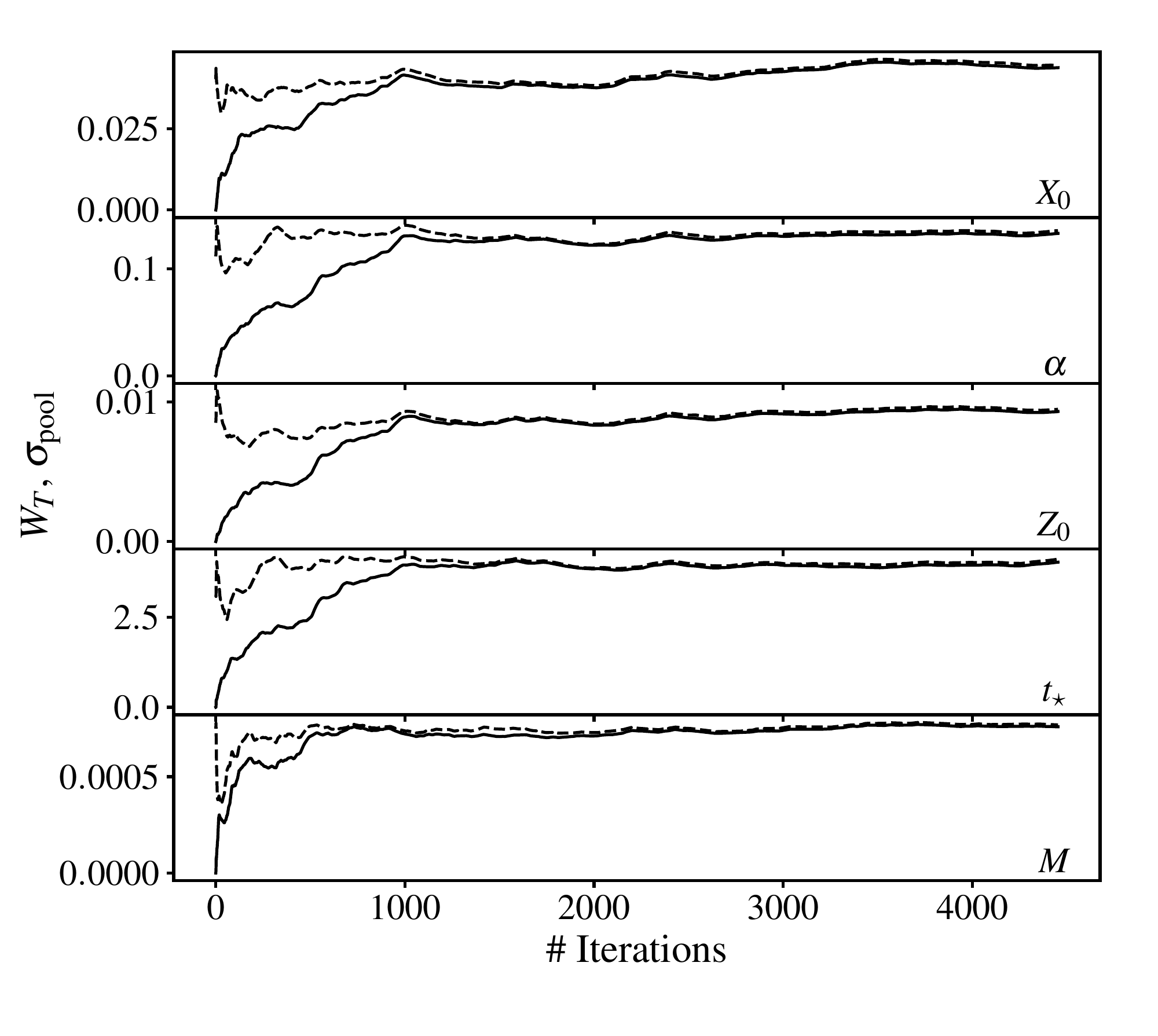}
\caption{Pooled (dashed lines) and within (full lines) variances for the stellar parameters.}
\label{fig:variances}
\end{figure}

\begin{figure}[t!]
\center
\includegraphics[width=0.475\textwidth]{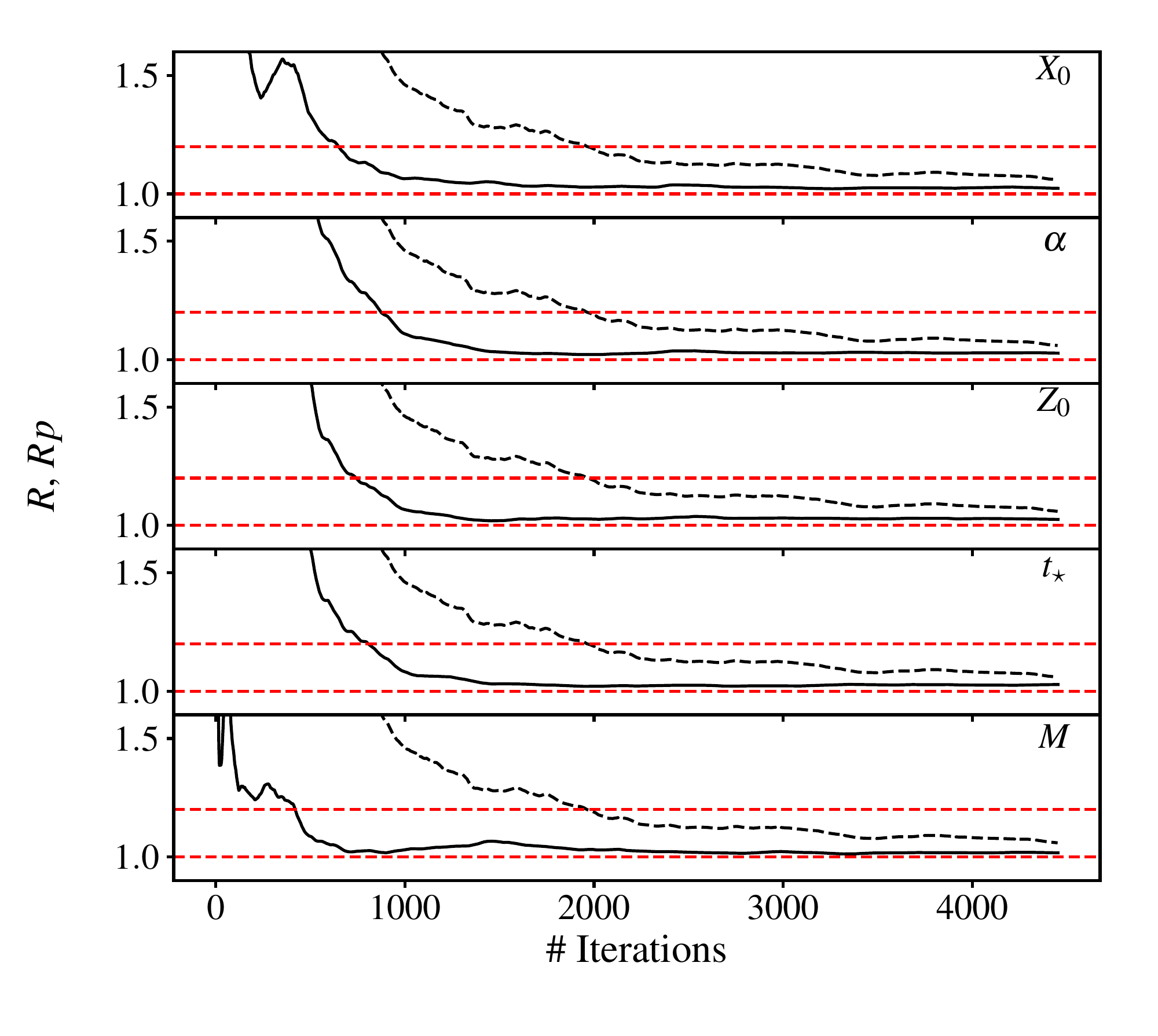}
\caption{Pseudo-scale reduction factor (full line) for all stellar parameters. The multivariate pseudo-scale reduction factor is shown in each panel as a black dashed line.}
\label{fig:criteria}
\end{figure}

\subsection{Convergence assessments}

We show here some common convergence diagnostics for a multiple chain sampler. In Fig.~\ref{fig:variances} we show the pooled and within variance and in Fig.~\ref{fig:criteria} the pseudo-scale reduction factor first introduced by \citet{Gelman92}. We use the proper correction for the degrees of freedom given in \citet{Brooks98}. We also show the posterior pooled and within variances. These three indicators taken together give a decent indication of convergence. First it appears that the two variances are fairly stabilised in all cases and converge towards each other (the within variance being lower as is expected). The pseudo-scale reduction factor is always lower than 1.2 which is a reasonable indicator for convergence (it is expected to converge to 1 when $N\rightarrow+\infty$).

These are indicators defined for univariate distributions. \citet{Brooks98} also provide a convergence diagnostic for multivariate distributions. We also display it in Fig.~\ref{fig:criteria}. It is an upper bound to the pseudo-scale reduction factor, as an MCMC algorithm converges slower towards the marginals than the joint posterior. Nevertheless it still decreases towards 1, giving another confirmation that our chains have converged towards the same stationary density. 

Finally, we also used the cumulative mean of the sample of these chains to control the mixing. These indicate that the algorithm may not be performing optimally. This is further confirmed by the fact that the acceptance rates are usually lower than what is expected. They vary between roughly 3\% and 10\%. We attribute this to the strong correlations seen in the PDF. 

\end{document}